\documentclass[12pt]{article}
\usepackage{epsfig,amssymb}
\usepackage{latexsym}

\newcommand{\bq}{\begin{equation}}
\newcommand{\eq}{\end{equation}}
\newcommand{\bqa}{\begin{eqnarray}}
\newcommand{\eqa}{\end{eqnarray}}
\newcommand{\ben}{\begin{enumerate}}
\newcommand{\een}{\end{enumerate}}
\newcommand{\bc}{\begin{center}}
\newcommand{\ec}{\end{center}}
\newcommand{\bqb}{\begin{eqnarray*}}
\newcommand{\eqb}{\end{eqnarray*}}

\def\pr#1#2#3{ Phys. Rev. ${\bf{#1}}$:#2 (#3)}
\def\prl#1#2#3{ Phys. Rev. Lett. ${\bf{#1}}$:#2 (#3)}
\def\pl#1#2#3{ Phys. Lett. ${\bf{#1}}$:#2 (#3)}
\def\prep#1#2#3{ Phys. Rep. ${\bf{#1}}$:#2 (#3)}
\def\rmp#1#2#3{ Rev. Mod. Phys. ${\bf{#1}}$:#2(#3)}

\def\np#1#2#3{ Nucl. Phys. ${\bf{#1}}$:#2 (#3)}

\def\jhep#1#2#3{ JHEP ${\bf{#1}}$:#2 (#3)}
\def\epj#1#2#3{ Eur. Phys. J. ${\bf{#1}}$:#2 (#3)}
\def\ijmp#1#2#3{ Int. J. Mod. Phys. ${\bf{#1}}$:#2 (#3)}

\def\aop#1#2#3{Annals of Phys. ${\bf{#1}}$:#2 (#3)}

\def\polon#1#2#3{Acta Phys. Polon. ${\bf{#1}}$:#2 (#3) }

\def\ie{{\it i.e. ~}}
\def\eg{{\it e.g. ~}}

\def\etal{{\it et.al.~}}

\global\nulldelimiterspace = 0pt

\def\sw{s_W}
\def\cw{c_W}

\def\mwd{m_W^2}
\def\mw{m_W}

\def\tchi{\tilde \chi}

\def\shat{\hat s}
\def\that{\hat t}
\def\uhat{\hat u}

\begin{document}
\pagenumbering{arabic}
\thispagestyle{empty}
\def\thefootnote{\fnsymbol{footnote}}
\setcounter{footnote}{1}

\begin{flushright}
June  2004, corrected version\\
PM/04-05\\
hep-ph/0404162\\

 \end{flushright}
\vspace{2cm}
%---------------------titre ---------------------------------------
\begin{center}
{\Large\bf Neutralino pair production at CERN LHC\footnote{Work supported
by the European Union under contract HPRN-CT-2000-00149.}}.
 \vspace{1.cm}  \\
%-----------------------------------------------------------------
{\large G.J. Gounaris$^a$, J. Layssac$^b$, P.I. Porfyriadis$^a$
and F.M. Renard$^b$}\\
\vspace{0.2cm}
$^a$Department of Theoretical Physics, Aristotle
University of Thessaloniki,\\
Gr-54124, Thessaloniki, Greece.\\
\vspace{0.2cm}
$^b$Physique
Math\'{e}matique et Th\'{e}orique,
UMR 5825\\
Universit\'{e} Montpellier II,
 F-34095 Montpellier Cedex 5.

\vspace*{1.cm}

{\bf Abstract}
\end{center}
\vspace*{-0.4cm}
We consider the production of neutralino pairs $\tchi^0_i\tchi^0_j$
at a high energy hadron collider, putting  a special emphasis on the
case where one of them is the lightest neutralino
$\tchi^0_1$, possibly constituting the main Dark Matter component.
At tree level, the only relevant subprocess is
$q\bar q\to \tchi^0_i\tchi^0_j$, while the subprocess
$gg\to \tchi^0_i\tchi^0_j$ first appears at the one loop level.
Explicit expressions for the $q\bar q$-helicity amplitudes are
presented, including the tree level contributions and
the leading-log one loop radiative corrections.
For the one-loop $ gg\to \tchi^0_i\tchi^0_j$ process,
a numerical code named PLATONggnn is released, allowing the
computation of $d\sigma /d \that$ in any MSSM model with real
soft breaking parameters.
It turns out that acceptable MSSM benchmark models exist for which
the $q\bar q $ and the gluonic  contributions
may give comparable effects at LHC,
due to the enhanced  gluonic structure functions at low
fractional momenta. Depending on the values of the MSSM parameters,
we find that the LHC neutralino pair production
may provide sensitive  tests of SUSY models generating neutralino
Dark Matter.

\vspace{0.5cm}

\def\thefootnote{\arabic{footnote}}
\setcounter{footnote}{0}
\clearpage

\section{Introduction}

Neutralino production at hadron colliders is an important part of the program of
 Supersymmetry (SUSY) searches \cite{SUSYsearches1,SUSYsearches2}.
One special reason is related to the
possibility that $\tchi^0_1$, the lightest neutralino state, is in fact
the Lightest Supersymmetric Particle (LSP) \cite{LSP}.
This has two particular consequences; the  first  concerning   the supersymmetric
spectroscopy (chains and rates of decays) in R-parity conserving
models \cite{SUSYsearches1}; while the  second
 largely determining    the search  for Dark Matter (DM)  \cite{DMLSP}.\par

DM detection  in such a case is expected
to occur either in a direct way (\eg  through the observation of nucleus recoil
in $\tchi^0_1 N \to \tchi^0_1 N$ elastic scattering); or in an
indirect way, by observing modifications of the cosmic spectrum of particles like
photons, positrons, antiprotons etc., due to contributions from
  $\tchi^0_1 \tchi^0_1$ annihilation   \cite{DMobs}.
Concerning the indirect way, we have presented in two previous papers
the  results of a complete  one-loop
computation for the processes $\tchi^0_i\tchi^0_j\to \gamma\gamma,gg$
involved in DM annihilation \cite{DMann}, as
well as the results for  the reversed process of
neutralino pair production at a photon-photon collider \cite{gammagammaDM}.
In \cite{DMann} we have also emphasized   that in certain benchmark
MSSM models the gluon-gluon channel may be
important for  determining the neutralino relic density
  \cite{Snowmass, Arnowitt, CDG}.

We would expect therefore, that for neutralino pair production   at a high energy
hadron collider like  LHC, kinematical domains may   exist
where   the gluon structure function of the proton is so
large \cite{MRST}, that the one-loop gluon annihilation contribution
 may in fact be bigger than the tree level $q\bar q$ contribution.
The precise study of such neutralino-pair production process at LHC, through
the subprocesses  $q\bar q \to \tchi^0_i\tchi^0_j$ and
$gg \to \tchi^0_i\tchi^0_j $,   constitutes the
aim of the present paper.\par

First, we present  the helicity amplitudes and cross sections of the
subprocess $q\bar q\to\tchi^0_i\tchi^0_j$ at
tree level. At this level, such a process has been studied long ago
\eg  in \cite{Dawson, Baer1}. We go beyond this though by
also exploring   the Fermi statistics and CP (for real MSSM parameters)
constraints on these amplitudes, which  strongly reduce their  number
and also serve as a check of the calculations.

In a second step, and in order to check the possible
existence of important one-loop electroweak (EW)
contributions to these $q\bar q$ amplitudes,
the leading ($\alpha  \ln^2s$) and
sub-leading ($\alpha  \ln s$)
 contributions are included, following
 the procedure established  in \cite{BRV, Denner,qqLHC}.
These  EW corrections
reduce the overall tree level magnitude of the
amplitudes by an amount that can reach the few tens of percent
level for the kinematical domain attainable
at LHC. In the same direction acts also the
SUSY QCD contribution\footnote{These involve QCD interactions explicitly
affecting SUSY particle exchanges \cite{qqLHC}.}
calculated according to the rules established
to  order $\alpha_s \ln s$ in  \cite{qqLHC, Beenakker}.
We should emphasize at this point though, that these EW and SUSY QCD
 corrections should be
considered in \underline{addition} to the pure QCD (leading and next-to-leading)
corrections which  strongly  increase the  tree level amplitudes, as
 found in  \cite{Beenakker}. \par

We then turn to  the one-loop subprocess $gg\to\tchi^0_i\tchi^0_j$,
for which the helicity amplitudes are calculated using
the set of diagrams established in  \cite{DMann, gammagammaDM}. There,   the
neutralino annihilation  amplitudes  $\tchi^0_i\tchi^0_j \to gg$ were
calculated  under  any kinematical conditions; but the accompanying
numerical codes compute the neutralino MSSM annihilation cross
section to gluons  only at the appropriate
for dark matter threshold region \cite{plato}.
Using these results, the  numerical code PLATONggnn has been also constructed,
which calculates the reversed process cross section
$d\sigma (gg  \to\tchi^0_i\tchi^0_j)/d\that $ for  any $(\shat,\that)$-values
and  any MSSM model with real soft breaking
parameters  \cite{plato}.

We then compute the LHC cross sections for $PP  \to\tchi^0_i\tchi^0_j+...$, by
convoluting the $gg$ and $q\bar q$ subprocess cross sections,
 with the corresponding quark
and gluon distribution functions in the initial protons $P$.
We then discuss the contributions of both
subprocess to several observables (invariant mass, transverse
momentum and angular distributions)
and we give illustrations for an extensive  set of
benchmark models in  MSSM.
As we will see below,  depending on the choice of MSSM
parameters  and  the kinematical regions looked at,
the one loop $gg\to \tchi^0_i\tchi^0_j$  subprocess   may
occasionally give comparable or even larger effects, than the tree level
$q\bar q\to\tchi^0_i\tchi^0_j$ one. \par

These results imply an interesting
complementarity between the future LHC measurements, the related
$\gamma \gamma \to \tchi^0_i\tchi^0_j$ measurements
 at  a future Linear Collider and the Dark Matter
searches  in cosmic experiments. \par

The contents of the paper is the following.
Sect.2 is devoted to the process $q\bar q\to\tchi^0_i\tchi^0_j$.
The general properties of the helicity amplitudes
are studied in the subsection 2.1,
where the seven basic independent amplitudes are identified.
The tree-level helicity amplitudes and cross sections are subsequently
presented in Section 2.2 and Appendix A.1;
while   the electroweak  and SUSY QCD corrections
to the helicity amplitudes,  at leading and subleading
logarithmic accuracy, are given  in Section 2.3 and Appendix A.2.
In Sect.3, the one loop process $gg \to\tchi^0_i\tchi^0_j$ is presented.
Applications to neutralino pair production at LHC
using the parton formalism are given in Sect.4, where the numerical
results are also discussed. The  concluding remarks are given in
Sect.5, while the parton model kinematics is detailed in
Appendix B.

\section{The subprocess $q\bar q\to\tchi^0_i\tchi^0_j$}

\subsection{Generalities about Helicity amplitudes for
$q\bar q \to \tchi^0_i\tchi^0_j$}

For an incoming $q\bar q$-pair, and an outgoing pair of neutralinos,
the process is written as
\bq
q (q_1, \lambda_1 ) ~\bar q(q_2, \lambda_2 ) \to
\tchi^0_i (p_i, \tau_i) ~\tchi^0_j(p_j, \tau_j)~~, \label{qq-process}
\eq
where $(q_1, q_2, p_i, p_j)$ and\footnote{The possible values of the
helicities $\lambda_1, \lambda_2, \tau_i,\tau_j$ are, as usually, taken as $\pm1/2$.}
$(\lambda_1, \lambda_2, \tau_i,\tau_j)$ are the momenta and
helicities of the incoming and outgoing particles.
The usual Mandelstam variables for the subprocess are defined as
\bq
\shat=(q_1+q_2)^2=(p_i+p_j)^2 ~,~~ \that=(q_1-p_i)^2=(q_2-p_j)^2~,~
\uhat=(q_1-p_j)^2=(q_2-p_i)^2  \label{stu-def}.
\eq

Since the  top quark structure function is vanishing in the proton
 and the other  quarks are not too heavy,
 the incoming $q$ and $\bar q$ in (\ref{qq-process})
are taken as \underline{massless}  as far as the kinematics is
concerned, but we keep the potentially large (particularly for the third family)
 Yukawa contributions to the couplings.  Finally
$(m_i,~m_j)$ denote  the $(\tchi^0_i,~\tchi^0_j)$
masses, respectively.

The helicity amplitude for  process (\ref{qq-process}) is denoted
as
\bq
F^{ij}_{\lambda_1\lambda_2;\tau_i\tau_j}(\theta^*)~~, \label{qq-Fij}
\eq
where  $\theta^*$  is    scattering angle in the c.m. of the subprocess.
In these  amplitudes,  $\bar q$ and $\tchi_j^0$ are treated
as particles No2 in the JW
 conventions \cite{JacobW}. To consistently take into account the Majorana
 nature of the neutralinos, we always describe the No1 neutralino $\tchi_i^0$
 through a positive energy Dirac wave function, while the No2 JW particle
  $\tchi_j^0$  is described through a negative energy
  one\footnote{The same convention is also followed for the
  $gg \to \tchi^0_i \tchi^0_j$ treated below.}.
Fermi statistics for the final neutralinos then
implies\footnote{We note that   (\ref{neutralino-Fermi-ex}), which is induced by
the anticommuting nature  of  the Fermionic fields, does not  generally agree
with the neutralino (anti)symmetry property  assumed in \cite{Liang}.}
\bq
F^{ij}_{\lambda_1\lambda_2;\tau_i\tau_j}(\theta^*)=(-1)^{\lambda_1-\lambda_2}
F^{ji}_{\lambda_1\lambda_2;\tau_j\tau_i}(\pi-\theta^*)~~,
\label{neutralino-Fermi-ex}
\eq
while CP-invariance, valid for real soft breaking and $\mu$ parameters, gives
\bq
F^{ij}_{\lambda_1\lambda_2;\tau_i\tau_j}(\theta^*)=\eta_i \eta_j
F^{ji}_{-\lambda_2,-\lambda_1;-\tau_j,-\tau_i}(\theta^*)~~, \label{CP-constraint}
\eq
where $\eta_i=\pm 1$ is  the CP-eigenvalue of the
$\tchi^0_i$-neutralino \cite{LeMouel}.

On the basis of (\ref{neutralino-Fermi-ex}, \ref{CP-constraint})
all $q\bar q$-amplitudes may be expressed in terms
of  \underline{seven basic ones}, selected as
\bqa
&& F^{ij}_{-+-+}(\theta^*)~~,~~F^{ij}_{+-+-}(\theta^*)~~,
~~F^{ij}_{-+++}(\theta^*)~~,~~F^{ij}_{+-++}(\theta^*)~~,~~
\nonumber \\
&& F^{ij}_{++++}(\theta^*)~~,~~F^{ij}_{++--}(\theta^*)~~,~~
F^{ij}_{+++-}(\theta^*)~~. \label{basic7}
\eqa
The other amplitudes are determined from these through
\bqa
F^{ij}_{-++-}(\theta^*)&=&- F^{ji}_{-+-+}(\pi-\theta^*)~~, \nonumber \\
F^{ij}_{+--+}(\theta^*)&=&- F^{ji}_{+-+-}(\pi-\theta^*)~~, \nonumber \\
F^{ij}_{-+--}(\theta^*)&=& F^{ji}_{-+++}(\theta^*)\eta_i\eta_j~~, \nonumber \\
F^{ij}_{+---}(\theta^*)&=& F^{ji}_{+-++}(\theta^*)\eta_i\eta_j~~, \nonumber \\
F^{ij}_{++-+}(\theta^*)&=& F^{ji}_{+++-}(\pi-\theta^*)~~, \nonumber \\
F^{ij}_{--++}(\theta^*)&=& F^{ji}_{++--}(\theta^*)\eta_i\eta_j~~, \nonumber \\
F^{ij}_{----}(\theta^*)&=& F^{ji}_{++++}(\theta^*)\eta_i\eta_j~~, \nonumber \\
F^{ij}_{--+-}(\theta^*)&=& F^{ji}_{+++-}(\theta^*)\eta_i\eta_j~~, \nonumber \\
F^{ij}_{---+}(\theta^*)&=&F^{ji}_{+++-}(\pi-\theta^*)\eta_i\eta_j~~. \label{rest9}
\eqa
We also note that  (\ref{neutralino-Fermi-ex}, \ref{CP-constraint})
 imply the  relations
\bqa
F^{ij}_{++++}(\theta^*)&=& F^{ji}_{++++}(\pi-\theta^*)~~, \nonumber \\
F^{ij}_{++--}(\theta^*)&=& F^{ji}_{++--}(\pi-\theta^*)~~, \nonumber \\
F^{ij}_{--++}(\theta^*)&=& F^{ji}_{--++}(\pi-\theta^*)~~, \nonumber \\
F^{ij}_{----}(\theta^*)&=& F^{ji}_{----}(\pi-\theta^*)~~, \nonumber \\
F^{ij}_{+-++}(\theta^*)&=& -F^{ji}_{+-++}(\pi-\theta^*)~~, \nonumber \\
F^{ij}_{-+++}(\theta^*)&=& -F^{ji}_{-+++}(\pi-\theta^*)~~, \nonumber \\
F^{ij}_{+---}(\theta^*)&=& -F^{ji}_{+---}(\pi-\theta^*)~~, \nonumber \\
F^{ij}_{-+--}(\theta^*)&=& -F^{ji}_{-+--}(\pi-\theta^*)~~. \nonumber \\
\eqa

In terms of these helicity amplitudes, the  unpolarized
differential subprocess cross section may expressed as
\bq
{d \hat \sigma (q\bar q \to \tchi^0_i\tchi^0_j)\over d\that}
= {1 \over 192 \pi \shat^2}
\Big ({1\over2}\Big )^{\delta_{ij}}
\Big [\sum_{\lambda_1,\lambda_2;\tau_i,\tau_j}
|F_{\lambda_1,\lambda_2;\tau_i,\tau_j}|^2 \Big ]~~,
\label{dsig-qq}
\eq
where the kinematics are defined in (\ref{stu-def}) and
(\ref{shat}-\ref{pstar}).

\subsection{Born amplitudes and cross sections.}

The Born amplitude  for the process in (\ref{qq-process})
contains  three diagrams
(see Fig.\ref{qq-diagrams}abc) involving  $s$, $t$ and $u$ channel
exchanges and written as \cite{Baer1}
\bq
F^{ijB}_{\lambda_1\lambda_2\tau_i\tau_j}
=F^{ijBs}_{\lambda_1\lambda_2\tau_i\tau_j}+
F^{ijBt}_{\lambda_1\lambda_2\tau_i\tau_j}+
F^{ijBu}_{\lambda_1\lambda_2\tau_i\tau_j} ~, \label{born}
\eq
\bqa
 F^{ijBs}_{\lambda_1\lambda_2\tau_i\tau_j}= -{e^2\over4s^2_Wc^2_W(\hat s-m^2_Z)}
 \bar v(\bar q)\gamma^{\mu}[g_{qL}P_L+g_{qR}P_R]u(q)\cdot
\bar u_i\gamma_{\mu}[N^{ij}P_L-N^{ij*}P_R]v_j, &&
\nonumber \\
 F^{ijBt}_{\lambda_1\lambda_2\tau_i\tau_j}=\sum_n {1\over \hat t-\tilde m^2_n}
\bar v(\bar q)[A^{L*}_j(\tilde q_n)P_R+A^{R*}_j(\tilde q_n)P_L]v_j\cdot
\bar u_i[A^{L}_i(\tilde q_n)P_L+A^{R}_i(\tilde q_n)P_R]u(q),&&
\nonumber \\
 F^{ijBu}_{\lambda_1\lambda_2\tau_i\tau_j}=-\sum_n {1\over \hat u-\tilde m^2_n}
\bar v(\bar q)[A^{L*}_i(\tilde q_n)P_R
+A^{R*}_i(\tilde q_n)P_L]u^c_i \cdot
\bar v^c_j[A^{L}_j(\tilde q_n)P_L+A^{R}_j(\tilde q_n)P_R]u(q),~~&&
 \label{bornstu}
\eqa
where the index $n$ refers to the summation over the exchanged
L- and  R- squarks  of the same flavor in the t- and u-channel,
$P_{L/R}=(1\mp \gamma_5)/2$,
and $(i,~j)$ describe the final neutralinos.

Explicit expressions for the seven basic helicity amplitudes listed in
(\ref{basic7}), are given in (\ref{Born-s-basic1}, \ref{Born-s-basic2},
\ref{Born-t-basic}, \ref{Born-u-basic}) in Appendix A.1.

\begin{itemize}

\item
They  involve  the  L and R $Zqq$-couplings  defined  as
\bq
 Zqq ~~ \Rightarrow ~   -{e\over2s_Wc_W} \gamma^{\mu}[g_{qL}P_L+g_{qR}P_R]~,
\label{Zqq1}
\eq
where
\bq
g_{qL}=2I^3_q(1-2s^2_W|Q_q|)~~~, ~~g_{qR}=-2s^2_WQ_q ~~, \label{Zqq2}
\eq
with  $I^3_q$, $Q_q$ being the isospin and charge of the
various $q_L$-quarks.

\item
The Z-neutralino couplings satisfying
\bq
Z\tchi^0_i\tchi^0_j ~~ \Rightarrow ~
 {e\over2s_Wc_W} \gamma^{\mu} [g_{ijL}P_L+g_{ijR}P_R] \label{Zchichi1}
\eq
with
\bq
 g_{ijL}= N^{ij}\equiv Z^{N*}_{4i}Z^{N}_{4j}-Z^{N*}_{3i}Z^{N}_{3j}~~,
 ~~ g_{ijR}=-N^{ij*}~, \label{Zchichi2}
\eq
where $Z^N$  denotes the neutralino mixing matrix  in the notation
of \cite{Rosiek}).

\item
And the neutralino-quark-squark couplings
\bq
q\tilde q_n \tchi^0_i  ~~\Rightarrow ~ ~
A^L_{i}(\tilde q_n)P_L+A^R_i (\tilde q_n )P_R ~, \label{qqchi1}
\eq
where
\bqa
  &A^{L}_{i}(\tilde{u}_L)=-{e\over 3\sqrt{2}s_Wc_W}(Z^N_{1i}s_W
+3Z^N_{2i}c_W) &,~
A^{L}_{i}(\tilde{u}_R)=-{e m_u\over \sqrt{2}M_Ws_W \sin\beta}Z^N_{4i}
\nonumber \\
 & A^{R}_{i}(\tilde{u}_R)={2e\sqrt{2}\over 3c_W}Z^{N*}_{1i}
&,~
A^{R}_{i}(\tilde{u}_L)=-{e m_u\over \sqrt{2}M_Ws_W \sin\beta}Z^{N*}_{4i}
\nonumber \\
& A^{L}_{i}(\tilde{d}_L)=-{e\over 3\sqrt{2}s_Wc_W}(Z^N_{1i}s_W
-3Z^N_{2i}c_W) &,~
A^{L}_{i}(\tilde{d}_R)=-{e m_d\over \sqrt{2}M_Ws_W \cos\beta}Z^N_{3i}
\nonumber \\
& A^{R}_{i}(\tilde{d}_R)=-{e\sqrt{2}\over 3c_W}Z^{N*}_{1i}
&,~
A^{R}_{i}(\tilde{d}_L)=
-{e m_d\over \sqrt{2}M_Ws_W \cos\beta}Z^{N*}_{3i}~. \label{qqchi2}
\eqa
\end{itemize}

In (\ref{qqchi2}),  $(q=u,~ d)$ refer to the incoming  up and down quark (antiquark) of
any family\footnote{As usual, we will only consider non-vanishing structure
functions for incoming u, d, s, c and b quarks.},
while $(\tilde q_n=\tilde q_L, ~\tilde q_R)$ denote the corresponding squarks.
We also note that the  mixing matrices $Z^N$ in (\ref{Zchichi2}, \ref{qqchi2}),
control  the Bino, Wino, Higgsino components of the neutralino
in the  $Z\tchi^0_i\tchi^0_j$ and $q\tilde q\tchi^0$ coupling
 \cite{Rosiek, LeMouel}.
 Finally, we remark that  $s$-channel Born part
 $F^{ijBs}_{\lambda_1\lambda_2\tau_i\tau_j}$ gives  non-vanishing
 contributions only  for  purely higgsino production,
 whereas the $t,u$-channel Born
parts for  purely gaugino.

One can  then compute the differential
cross section either through (\ref{dsig-qq}) and the helicity
amplitudes in (\ref{Born-s-basic1}, \ref{Born-s-basic2}, \ref{Born-t-basic},
\ref{Born-u-basic}),   or directly by the trace procedure
giving
\bq
{d \hat \sigma (q\bar q \to \tchi^0_i\tchi^0_j)\over d\that}
= {1 \over 192 \pi \shat^2}
\Big ({1\over2}\Big )^{\delta_{ij}}
[I_{ss}+I_{tt}+I_{uu} -2I_{st}+2I_{su}-2I_{tu}]
\label{dsig-qq1}
\eq
where\footnote{Analogous expression for gaugino production
in $e^-e^+$ Colliders have appeared in  \cite{epem}. }
\bqa
I_{ss}&=&{e^4(g^2_{qL}+g^2_{qR})\over4s^4_Wc^4_W(\shat -m^2_Z)^2}
\Big \{N^{ij}N^{ij*}[(m^2_i-\hat t)(m^2_j-\hat t)+(m^2_i-\hat u)(m^2_j-\hat u)]
\nonumber\\
&& -[(N^{ij})^2+(N^{ij*})^2]m_im_j\shat  \Big \}~, \nonumber \\
I_{tt}&=&\sum_{k,l}{(m^2_i-\hat t)(m^2_j-\hat t)
\over(\hat t-\tilde m^2_k)(\hat t-\tilde m^2_l)}[A^L_i(\tilde q_k)A^{L*}_i(\tilde q_l)
+A^{R}_i (\tilde q_k)A^{R*}_i(\tilde q_l) ][A^{L}_j(\tilde q_k)A^{L*}_j(\tilde q_l)
\nonumber \\
&& +A^{R}_j(\tilde q_k)A^{R*}_j(\tilde q_l) ]~, \nonumber \\
I_{uu}&=&\sum_{k,l}{(m^2_i-\hat u)(m^2_j-\hat u)
\over(\hat u-\tilde m^2_k)(\hat u-\tilde m^2_l)}[A^{L*}_i(\tilde q_k)
A^{L}_i(\tilde q_l)
 +A^{R*}_i(\tilde q_k) A^{R}_i(\tilde q_l)][A^{L*}_j(\tilde q_k)A^{L}_j(\tilde q_l)
\nonumber \\
&& +A^{R*}_j(\tilde q_k)A^{R}_j(\tilde q_l)] ~, \nonumber \\
I_{st}&=&\sum_{k}{e^2
\over2s^2_Wc^2_W(\hat t-\tilde m^2_k)(\shat -m^2_Z)} Re \Big \{
[N^{ij*}g_{qR}A^{R}_i(\tilde q_k)A^{R*}_j(\tilde q_k)
 - N^{ij}g_{qL}A^{L}_i(\tilde q_k)A^{L*}_j(\tilde q_k)]
 \nonumber \\
&&  \cdot (m^2_i-\hat t)(m^2_j-\hat t)
  +[N^{ij*}g_{qL}A^{L}_i(\tilde q_k)A^{L*}_j(\tilde q_k)-
N^{ij}g_{qR}A^{R}_i(\tilde q_k)A^{R*}_j(\tilde q_k)]m_i m_j\shat  \Big \}
~, \nonumber \\
I_{su}&=&\sum_{k}{e^2
\over2s^2_Wc^2_W(\hat u-\tilde m^2_k)(\shat -m^2_Z)}Re \Big \{
[N^{ij*}g_{qL}A^{L*}_i(\tilde q_k)A^{L}_j(\tilde q_k)
-N^{ij}g_{qR}A^{R*}_i(\tilde q_k)A^{R}_j(\tilde q_k)]
\nonumber \\
&& \cdot (m^2_i-\hat u)(m^2_j-\hat u)
+[N^{ij*}g_{qR}A^{R*}_i(\tilde q_k)A^{R}_j(\tilde q_k)-
N^{ij}g_{qL}A^{L*}_i(\tilde q_k)A^{L}_j(\tilde q_k)]m_im_j \shat  \Big \}
\nonumber \\
I_{tu}&=&\sum_{k,l}{1
\over(\hat u-\tilde m^2_l)(\hat t-\tilde m^2_k)}Re \Big \{
[A^{L}_i(\tilde q_k)A^{L}_i(\tilde q_l)A^{L*}_j(\tilde q_k)A^{L*}_j(\tilde q_l)
\nonumber \\
&& + A^{R}_i(\tilde q_k)A^{R}_i(\tilde q_l)A^{R*}_j(\tilde q_k)A^{R*}_j(\tilde q_l)]
m_im_j \shat
 +{1\over2}\Big [A^{L*}_i(\tilde q_k)A^{R*}_i(\tilde q_l)A^{R}_j(\tilde q_k)
 A^{L}_j(\tilde q_l)
 \nonumber \\
&& + A^{R*}_i(\tilde q_k)A^{L*}_i(\tilde q_l)A^{L}_j(\tilde q_k)A^{R}_j(\tilde q_l)
\Big ] [(m^2_i-\hat t)(m^2_j-\hat t)+(m^2_i-\hat u)(m^2_j-\hat u)
\nonumber \\
&&  -\shat (\shat -m^2_i-m^2_j)] \Big \}~, \label{dsig-qq2}
\eqa
where $m_i,~ m_j$ are the neutralino masses and $N^{ij}$ are defined
in (\ref{Zchichi2}).

The  results in eq.(\ref{dsig-qq1}) disagree with
those of  \cite{Liang}, where  symmetry
properties of the $\tchi^0_i\tchi^0_j$
states which are different from those in
(\ref{neutralino-Fermi-ex}) have been used.\\

\subsection{One loop electroweak and SUSY QCD
corrections\\ to $q\bar q\to\tchi^0_i\tchi^0_j$}

In principle, one loop EW corrections  for $q\bar q\to\tchi^0_i\tchi^0_j$
should be taken into account, particularly because the energy reach at LHC
is so big, that the large logarithmic contributions to the amplitudes  may reach
 the few tens of percent level \cite{BRV, Denner, qqLHC}.
 In  the models we have considered,
 this implies  a reduction of the cross sections sometimes by  almost
 a factor of two, while preserving their shape.
 Since the non-logarithmic  one-loop contributions
 seem to lie at the few percent level, which is also the level of
the expected experimental   accuracy, it  may be
adequate  to  ignore these difficult to calculate effects
in  $(q\bar q\to\tchi^0_i\tchi^0_j)$ at LHC
energies.

In this section we present therefore, the leading and subleading
EW logarithmic contributions to the
$q\bar q\to \tchi^0_i\tchi^0_j$ helicity amplitudes,
following  \cite{BRV, Denner, qqLHC},
where applications for LC and LHC have been given.
They  are separated into three types
of   terms which are:
\begin{itemize}

\item \underline{Universal electroweak (EW) terms}.
These are process-independent terms appearing as correction
factors to the Born amplitude. They consist of "gauge" and "Yukawa"
contributions   associated to each external line and
determined  by its quantum numbers and chirality. Their
expressions for a quark or neutralino line are
respectively determined as follows:
\subitem \underline{External quark  line of chirality $a=L,~R$}:
Since all quarks are taken as massless as far as  the
kinematics are concerned\footnote{The  large third
family Yukawa terms appear only as couplings and do not
concern the kinematics.}, the quark lines correspond
to a definite chirality $a$. The induced  correction then    is
\bq
F^{ij B}_{a} \cdot [c(q)_a ] ~~, \label{cq-LL}
\eq
where $F^{ij B}_{a}$ describes the corresponding Born
amplitude from (\ref{born}) involving an external quark
line of chirality $a$ (which at high energies is essentially
equivalent  to the helicity),
while $(i,j)$-count the mass eigenstates of the neutralinos.
The coefficient in (\ref{cq-LL}) is written as
\bq
c(q)_a=c(q, ~{ \rm gauge})_a~+~c(q,~{\rm yuk})_a ~~,\label{cq}
\eq
with  the gauge contribution  being
\bq
c(q, ~{\rm gauge})_a={\alpha\over8\pi}\Big [{I_q(I_q+1)\over s^2_W}
~+~{Y^2_q\over4c^2_W}\Big ]
\Big (2 \ln{\shat \over m^2_W}-\ln^2{\shat \over m^2_W}\Big )~,
~ \label{cq-gauge}
\eq
where $I_q$ is the full isospin of the quark  $q_a$
with  chirality $a=L$ or $R$, and  $Y_q=2(Q_q-I^{(3)}_q)$ defines its
hypercharge in terms of the third isospin component $I^{(3)}_q$.
Correspondingly, the  Yukawa term (for $b,~t$ quarks only) is
\bqa
c(q,~{\rm yuk})_a&=&- {\alpha\over16\pi s^2_W}\Big [ln{\shat \over m^2_W}\Big ]
\Big \{ \Big
[{m^2_t\over m^2_W \sin^2\beta }+{m^2_b\over m^2_W\cos^2\beta }\Big ]\delta_{aL}
\nonumber\\
&&+2\Big [{m^2_b\over m^2_W \cos^2 \beta}~ \delta_{I^{(3)}_{q},-1/2}
+{m^2_t\over m^2_W \sin^2 \beta}~ \delta_{I^{(3)}_{q},1/2}\Big ]\delta_{aR}\Big \}~.
\label{cq-yuk}
\eqa

We note that an external antiquark
line should be counted separately giving  an additional
contribution determined by the same formulae
(\ref{cq-LL}-\ref{cq-yuk}).
Moreover,  the same  formulae describe also the
logarithmic contributions associated with each
external squark, anti-squark, lepton or
slepton line  \cite{BRV, Denner, qqLHC}.

\subitem
\underline{For an  external neutralino  line of chirality $b$},
it is convenient to  use a matrix notation
\bq
\sum_k \Big [F^{ikB}_{b} c(\tchi^0_k\tchi^0_j)_b
+F^{kjB}_{b} c^*(\tchi^0_k\tchi^0_i)_b \Big ] ~, \label{cchi-LL}
\eq
and to separate the higgsino from the gaugino components of the matrix
elements:
\bqa
c(\tchi^0_l\tchi^0_{l'})_b&=&
c(\tchi^0_l\tchi^0_{l'}~{\rm higgsino,gauge})_b
+ c(\tchi^0_l\tchi^0_{l'}~{\rm higgsino,yuk})_b\nonumber\\
&+ & c(\tchi^0_l\tchi^0_{l'}~{\rm gaugino,gauge})_b ~~, \label{cchi}
\eqa
with
\bqa
&& c(\tchi^0_l\tchi^0_{l'}~\rm{higgsino,gauge})_b=
{\alpha(1+2c^2_W)\over32\pi s^2_Wc^2_W}
\Big (2 \ln{\shat\over m^2_W}-\ln^2{\shat \over m^2_W}\Big )\nonumber\\
&& ~~~ \cdot  \Big [(Z^{N*}_{4l}Z^{N}_{4l'}+Z^{N*}_{3l}Z^{N}_{3l'})\delta_{bL}
+(Z^{N}_{4l}Z^{N*}_{4l'}+Z^{N}_{3l}Z^{N*}_{3l'})\delta_{bR}\Big ] ~,
\nonumber \\
&& c(\tchi^0_l\tchi^0_{l'}~\rm{higgsino,yuk})_b=
- {3\alpha\over16\pi s^2_W \mwd} \Big [\ln{\shat \over m^2_W}\Big ]
\nonumber \\
&& ~~~\cdot
\Big [{m^2_t\over  \sin^2 \beta}(Z^{N*}_{4l}Z^{N}_{4l'}\delta_{bL}
+Z^{N}_{4l}Z^{N*}_{4l'}\delta_{bR})
 +{m^2_b\over  \cos^2 \beta}(Z^{N*}_{3l}Z^{N}_{3l'}\delta_{bL}
+Z^{N}_{3l}Z^{N*}_{3l'}\delta_{bR}) \Big ]~, \nonumber \\
&& c(\tchi^0_l\tchi^0_{l'}
~{\rm gaugino,gauge})_b
=- {\alpha \over4\pi s^2_W}\Big [Z^{N*}_{2l}Z^{N}_{2l'}P_L+
Z^{N}_{2l}Z^{N*}_{2l'}P_R \Big ]
\Big [\ln^2{\shat \over m^2_W}\Big ] \label{cchi1}~.
\eqa

The logarithmic contributions associated to the purely higgsino
s-channel Born amplitude $F^{ikBs}_b$
of\footnote{See also (\ref{born}).} Fig.\ref{qq-diagrams}a  only involve
the "higgsino, gauge" and "higgsino, Yukawa" elements,
whereas the contributions associated to the purely gaugino
t- and u- Born amplitudes $F^{ikBt,u}_b$,
only involve the "gaugino, gauge" elements.

\item
 \underline{Angular and process dependent terms}. They originate
from  diagrams involving
$W$ internal lines supplying  soft-infrared $\ln^2t$ or $\ln^2u$ terms.
Diagrams with internal $Z$ lines are negligible,
since their contributions turn out to be  orthogonal
to the Born terms  and cannot interfere with them.
 The contributing diagrams therefore consist of
boxes with an intermediate $WW$ pair
in the s-channel,  triangles involving a single $W$
connected to a squark exchange in the t or
u channels, and   boxes involving  a single $W$
and a squark in the t or  u channels.

\item
 \underline{Renormalization Group (RG) terms}.
They arise from  intermediate $Z$ boson
Born terms contributing to   Higgsino production
only and inducing   running effects to
the  corresponding  $g$ and $g'$ gauge couplings; see Fig.1a.
In terms of the s-channel Born amplitudes of  (\ref{born}), they
are written as
\bq
F^{RG}=-{1\over4\pi^2}\Bigg (g^4\tilde\beta_0 {dF^{ijBs}\over dg^2}
+g'^4\tilde\beta'_0 {dF^{ijBs}\over dg'^2}\Bigg )
\Big [\ln{\shat \over \mu^2}\Big ]~,
\eq
where $g^2=e^2/s^2_W,~g'^2=e^2/c^2_W$,
\bqa
{\tilde \beta}_0= \frac{3}{4} C_A- \frac{n_g}{2}-\frac{n_h}{8}
=-~{1\over4}~~&,&~~
 {\tilde \beta}_0^\prime=-\frac{5}{6}n_g-\frac{n_h}{8}
=-~{11\over4}
\eqa
and   $C_A=2$, $n_g=3$, $n_h=2$ in MSSM.
Applying  this procedure to the 7 basic Born helicity amplitudes
of Appendix A.1, implies the substitutions
\bqa
{e^2g_{qL}\over s^2_Wc^2_W} & \to &
-~{(2I^{(3)}_{qL})\over4\pi^2}
\Big [\ln{\shat \over \mu^2}\Big ]
\{\tilde\beta_0g^4+\tilde\beta'_0g'^4[1-2|Q_q|]\}~, \nonumber \\
{e^2g_{qR}\over s^2_Wc^2_W} & \to &
~{(2Q_q)\over4\pi^2}
\Big[\ln{\shat \over \mu^2}\Big ]
\{\tilde\beta'_0g'^4\}~,
\eqa
\end{itemize}

The resulting expression for the seven basic helicity amplitudes
of (\ref{basic7}), are given  in Appendix A.2.
For what concerns the magnitude of the various
corrections, the following general comments can be made.
In the LHC domain (say for $\sqrt{\hat s}\simeq 1$ TeV)
a single $\ln (\hat s/ \mwd)  $
and a squared $\ln^2 (\hat s/\mwd) $ give  enhancement factors
of about $5$ and $25$, respectively. So one expects that
the corrections are of the order of $-5\%$ for the
quark or higgsino-gauge terms, $-25\%$ for the gaugino gauge ones, and
$-10\%$ for the  higgsino-Yukawa terms (depending on $\tan\beta$ value).
The angular dependent terms have a more complicated structure; their
sign is not fixed, while  their magnitude  can  reach the $10\%$ level.
The addition of   these various electroweak terms is strongly
model dependent, especially due to  the $Z^N_{ij}$
matrix elements  controlling  the amount of the higgsino and
 gaugino components of the neutralinos.
The net EW effect  on the amplitude though, is essentially always
negative and can easily reach the several tens of percent level.\\

\underline{The logarithmic SUSY QCD} corrections
for quark-antiquark processes  at order $\alpha_s$
are given by    \cite{Beenakker, qqLHC},
\bq
F^{SUSY~QCD}_{\lambda_q,\lambda_{\bar q};\tau_i,\tau_j}
=F^{Born}_{\lambda_q,\lambda_{\bar q};\tau_i,\tau_j}
\Big [-{\alpha_s\over3\pi}\ln{\hat s\over M^2  }\Big ] ~.
\eq
This simple logarithmic terms  arise  from  diagrams
containing virtual squarks and gluinos interacting via SUSY QCD
couplings; such diagrams produce no $\ln^2 s$-terms.
By itself this single log term is of course negative,
and it would remain around  $-5\%$ in the observable LHC domain.
In addition to these though, we should always also
consider the pure QCD leading and next-to-leading
mass dependent corrections which, as shown in ref.\cite{Beenakker},
 result into a K factor that turns out to be positive
and of the order of 30\%.  These later corrections are
not  discussed in the present paper.

\section{The one loop process $gg\to\tchi^0_i\tchi^0_j$}

This process first appears
at the one loop level through the triangle and box diagrams
fully    listed in \cite{DMann}.
These diagrams basically involve gluon-quark-squark
and neutralino-chargino-squark couplings;
no gluino can appear at this order. Accidental
degeneracies between the neutralino masses and  squark masses
can give some enhancement effects.
In addition,  single $Z,~h^0,~H^0$ or $A^0$ exchanges
in the s-channel,
can also give enhancements and resonance effects at the
corresponding c.m. energies. These situations are rather similar
to those already mentioned for the $\gamma\gamma\to \tchi^0_i\tchi^0_j$
process in \cite{gammagammaDM}.

The helicity amplitudes for  the process
\bq
 g(q_1,\mu_1) g(q_2,\mu_2) \to \tchi^0_i(p_i, \tau_i)
 \tchi^0_j(p_j,\tau_j)~, \label{gg-process}
\eq
are denoted as
\bq
F^{ij}_{\mu_1\mu_2;\tau_i\tau_j}(\theta^*)~~, \label{gg-Fij}
\eq
where the momenta and helicities of the incoming gluons and
outgoing neutralinos are  defined, and $\theta^*$ again
denotes the c.m. scattering angle.
We use the same $(\tchi^0_i,~ \tchi^0_j)$ conventions as in
the $q\bar q $ case
and  in \cite{gammagammaDM},  implying
\bq
  F^{ij}_{\mu_1,\mu_2;\tau_i,\tau_j}(\theta^*)=(-1)^{\mu_1-\mu_2}~
 F^{ji}_{\mu_1,\mu_2;\tau_j,\tau_i}(\pi-\theta^*) ~~
\label{fermion-antisymmetry}
\eq
from $\tchi^0_i\tchi^0_j$ fermion-antisymmetry, and
 \bq
 F^{ij}_{\mu_1,\mu_2;\tau_i,\tau_j}(\theta^*)
=(-1)^{\tau_i-\tau_j}~
 F^{ij}_{\mu_2,\mu_1;\tau_i,\tau_j}(\pi-\theta^*)~~
\label{boson-symmetry}
\eq
from $gg$-boson symmetry.

If the MSSM breaking parameters and  the Higgs parameter
$\mu$ are real, then CP invariance holds, implying
\bq
  F^{ij}_{-\mu_1,-\mu_2;-\tau_i ,- \tau_j}(\theta^*)=
(-1)^{\tau_i-\tau_j-(\mu_1-\mu_2)}~\eta_i\eta_j
 F^{ij}_{\mu_1,\mu_2;\tau_i,\tau_j}(\theta^*) ~~,
\label{CP-invariance}
\eq
where $\eta_i,~ \eta_j=\pm 1$ are  the CP-eigenvalues of the two produced
neutralinos\footnote{We follow the same notation as in \eg \cite{LeMouel}.}.
In such a case, time inversion invariance implies
 the same helicity amplitudes for the process
(\ref{gg-process}) and its inverse.
Combining  (\ref{fermion-antisymmetry},
\ref{boson-symmetry}, \ref{CP-invariance}), we get
\bq
 F^{ij}_{\mu_1\mu_2; \tau_i\tau_j}(\theta^*)=
(-1)^{\mu_1-\mu_2 +\tau_j-\tau_i}
 F^{ji}_{\mu_2\mu_1; \tau_j\tau_i}(\theta^*)=
\eta_i \eta_j \tilde F^{ji}_{-\mu_2, -\mu_1;-\tau_j,-\tau_i}(\theta^*)~~,
\label{neutralino-photon-symmetry}
\eq
where the first part comes from (\ref{fermion-antisymmetry},
\ref{boson-symmetry}) alone, while for the last part the CP-invariance
relation (\ref{CP-invariance}) is also used.

In terms of these helicity amplitudes, the unpolarized differential
subprocess cross section is
\bq
{d \hat \sigma(gg \to \tchi_i \tchi_j )
\over d\that }=
{1 \over 4096 \pi \hat s^2 } ~\Big({1\over2}\Big )^{\delta_{ij}}
\sum_{\mu_1 \mu_2 \tau_i \tau_j} | F_{\mu_1\mu_2\tau_i\tau_j}|^2 . \label{dsig-gg}
\eq
Together with the present paper, we release in   \cite{plato}
  the numerical code PLATONggnn, which
calculates  the differential cross section  (\ref{dsig-gg})  as a functions of
$\theta^*$ and $\shat$, for any
set of real SUSY parameters at the electroweak scale.\\

The above amplitudes are basically of order $\alpha_s/\pi$
weaker than the tree level $q\bar q$ amplitudes of the preceding
Section. In certain SUSY models though, this reduction
can be partially compensated by the aforementioned
enhancement factors. But in practice, the most important
feature at LHC is the relative size of $gg$ and $q\bar q$ distribution
functions inside the proton; \ie the fact that at
"low" subenergies the  $gg$ fluxes are
much larger than $q\bar q$ ones. Because of this, and
as we  see in Section 4,
there are benchmark models where the $gg$ contribution to
neutralino-neutralino production at LHC, is
larger than the $q\bar q$ one.

\section{$\tchi^0_i\tchi^0_j$ distributions  in Proton-Proton collisions}

In this section we discuss numerical results for the
process $PP\to \tchi^0_i\tchi^0_j+....$ at LHC
(c.m. energy $\sqrt{s}=14$ TeV) generated  by the
subprocesses $gg, q\bar q\to \tchi^0_i\tchi^0_j$; ($\tchi^0_i$ is
always taken heavier than $\tchi^0_j$). All necessary
formulae describing the kinematics for two massive final particles
are presented in Appendix B. These allow the computation of
the neutralino-pair invariant mass  distributions $d\sigma/d\shat$,
the transverse energy distribution of the heavier neutralino
$d\sigma/ dx_{Ti}$, and  the angular distribution in
the neutralino-neutralino center
of mass  $d\sigma/ d\chi_i$
described through $\chi_i$ defined in (\ref{chi});
compare the formalisms in
Sections B.2.3, B.2.1 and B.2.4 respectively.

The present  formalism allows of course to compute any $\tchi^0_i\tchi^0_j$ channel
for any MSSM model with real soft breaking and $\mu$ parameters, using
 code PLATONggnn released in  \cite{plato}. In this
paper we also assumed that the $\tchi^0_1$  escapes the detector
without being observed\footnote{This would be the case if $\tchi^0_1$
is the stable LSP.}, so that the identification of \eg
the  $\tchi^0_2\tchi^0_1$-production
is only done through the detection of $\tchi^0_2$.
Our illustrations  are  restricted to the
$\tchi^0_2\tchi^0_1$ and $\tchi^0_2\tchi^0_2$ channels.

 As an example, for the quark and gluon
distribution functions inside the proton  we use the MRST2003c
package \cite{MRST} at the scale (\ref{Q-scale}).
Using this and the   31 benchmarks models \cite{Snowmass, Arnowitt, CDG}
already considered in the previous papers\footnote{The high scale values
of the defining parameters of these models are listed in Tables 1,2,3 of \cite{DMann}.}
\cite{DMann, gammagammaDM}, we have made numerical computations of
the above three single variable distributions.
Two  main features come out from this  study.

At high invariant subprocess energies (close to 1 TeV), the $gg$ contribution
becomes negligible compared to the $q\bar q$-ones.
This is due to  two effects; the $gg$ subprocess cross section is reduced by
the one loop factor $\alpha_s/\pi$ compared to the tree level
$q\bar q$-subprocess, while   the $gg$ luminosity is comparable to
(or even weaker  than)  the $q\bar q$ one.

On the opposite energy site,  within a few hundreds
of GeV above  threshold, the $gg$ flux may be  so large, that the
$gg$  contribution may compete
or  even overpass  the $q\bar q$-contribution
by a factor of  10 or more in some of the benchmark models.
This is further enhanced in cases  where the  $A^0$ or $H^0$ Higgs boson
can couple to  $\tchi^0_2\tchi^0_1$ or $\tchi^0_2\tchi^0_2$ channels.
\par

For the illustrations presented here, we have mainly selected
those of the benchmark models mentioned   above
where  the competition between the  $gg$ and $q\bar q$ contributions
is  most spectacular.
Thus, in Figs.\ref{SPS1a-fig}-\ref{SPS8-fig} we show the
invariant mass distribution $d\sigma/ d\shat $ in the
cases where there is a resonant enhancement
in the $gg$ contribution (models
SPS1a1, SPS5, SPS6, SPS8 ,\cite{Snowmass}), while
in Figs.\ref{SPS7-fig}, \ref{CDG24-fig}  the $gg$ contribution
is smooth but  important at low subenergies
(models SPS7, CDG24, \cite{Snowmass, CDG}), and it  is further reduced
in\footnote{The $gg$ contribution may become even smaller in
 some of the other   benchmark models mentioned above.} SPS4 (Fig.\ref{SPS4-fig}).

In all  cases, one sees that the $gg$ contribution
to $d\sigma/d\shat $ has a larger slope than the $q\bar q$
contribution; the effect being mainly due to the
behavior of the gluon distribution functions.
The precise magnitude of the $gg$ contribution
is however strongly  model dependent and arises as the result of
many features of the SUSY spectrum involved in the one loop diagrams
contributing to $gg\to\tchi^0_i\tchi^0_j$ \cite{DMann}.\par

As already discussed at the end of
Sect.2.3, the one loop logarithmic electroweak corrections  to
$q \bar q \to \tchi^0_i \tchi^0_j$
tend to reduce the size of the $q\bar q$ tree level cross sections;
see also \cite{BRV, Denner, qqLHC}.
They are strongly model dependent (due to the
neutralino mixing matrix  and  $\tan\beta$),
and can reach several tens  percent for the amplitudes.
For certain models, the addition
of the various terms can lead to a reduction of the size of the
$q\bar q$ cross sections by almost a factor two, but they do not
strongly modify their shapes. This reduction will of course
be somewhat reduced by the enhancements induced by the pure QCD effects
\cite{Beenakker}, but it is nevertheless sufficient
 to increase  the relative
importance  of the $gg$ contribution.

The features  observed in the  $d\sigma/ d\shat $
distributions of Figs.\ref{SPS1a-fig}-\ref{SPS4-fig},
can also be seen in the transverse energy distributions
$d\sigma/ dx_{Ti}$.  In order to not
multiply the number of figures, we only give illustrations
of this fact for two typical models,  SPS1a in Fig.9
(where there is a resonance), and  SPS7 in
Fig.10 (where there is no resonance).

Finally we have examined the distribution
$d\sigma/  d\chi_i$, which is essentially controlled
by the neutralino-neutralino center of mass
angular distribution; \ie the  $\cos\theta^*$ dependence discussed
in Appendix B2.4 and (\ref{chi}).
Such  $\chi_i$ distributions  can also be used as a
complementary test of the dynamics responsible for
neutralino-neutralino production. Typical illustrations
for models SPS1a  and SPS7 appear in Fig.11 and  Fig.12 respectively.
Concerning them we should remark, that the  EW correction to the
$q\bar q$ subprocess are reliable  mainly in large
$\chi_i$ region, where
 the $gg$ subprocess may also be important, especially
if there is an $A^0$ or $H^0$ resonance effect.
This appears as a threshold
effect corresponding to the value of $\chi_i$ above which
the mass of the resonance lies within the allowed
integration domain for (\ref{chii}).
On the contrary, the large correction
in the small  $\chi_i$ range of Figs.11,12
should not be taken too seriously, since
it is caused from a region where
$|t|$ is  small   and the leading-log predictions
not  valid.

\section{Final discussion}

In this paper we have considered the neutralino pair production
processes in proton-proton collisions at LHC.
In the description we have taken into account
the subprocess $q\bar q\to \tchi^0_i\tchi^0_j$ calculated
at the Born level as a first option, and as a second option we have  included  also
 leading and subleading logarithmic corrections. The  genuine one loop
$gg\to \tchi^0_i\tchi^0_j$ subprocess, is  fully taken into account.
The description applies to any MSSM model with real soft breaking and
 $\mu$ parameters. Analytic expressions for
the helicity amplitudes have been explicitly written for $q\bar q$-subprocess,
while a numerical code PLATONggnn is released allowing the computation of
the rather involved $d\hat \sigma(gg\to \tchi^0_i\tchi^0_j)/d\that$,
for any neutralino pair \cite{plato}.

After convoluting the subprocess
cross sections with parton distribution functions,
several observable  distributions in
$PP\to \tchi^0_i\tchi^0_j+....$  at LHC, have been studied.
For the applications, we have restricted   to $\tchi^0_2\tchi^0_1$ and
$\tchi^0_2\tchi^0_2$ production in the context of  31 benchmarks MSSM
models also considered in \cite{DMann, gammagammaDM}.
A strong model dependence is  observed, to which
  almost all aspects of the MSSM spectrum contribute
through masses and mixing matrix elements.

One of the most striking feature  we have found
is the important role of the $gg$ subprocesses, which, although
basically suppressed by the one loop $\alpha_s/\pi$ factor,
it may  occasionally  supply  a larger contribution than the
$q\bar q$ subprocess.
This  may occur  close to and slightly above threshold,
where the $gg$ luminosity could be sufficiently  large
to compensate for the  $\alpha_s/\pi$ factor.
In some models, the one loop $gg$ amplitudes may be  further enhanced
by the presence of $A^0$ or $H^0$
Higgs boson resonances, and possibly also by accidental
degeneracies between  the neutralino and  squark masses.

We have also given detail  illustrations for the  invariant mass,
transverse energy and angular distributions,
for the case  of seven  benchmark models where the $gg$ contribution
is generally spectacular \cite{Snowmass, Arnowitt, CDG};
\ie appearance of peaks, threshold effects etc.
In the few hundred  GeV subenergy domain, such structures
of  the $d\sigma/d\shat $  distributions, appear
in the range of 1 to 100 ${\rm fb/TeV^{-2}}$ and should be
observable at LHC. This may  also be true for the $d\sigma/dx_{Ti}$
and $d\sigma/d\chi_i$ distributions; compare Figs.9-11.
We should remember though, that there exist
benchmark models also, where the  $gg$ contribution
is rather marginal, as \eg in model
SPS4 \cite{Snowmass} and others \cite{Arnowitt, CDG}.

In all cases, the effect of the one loop logarithmic
corrections to the $q\bar q$ cross sections  appear to be
at the few tens of percent level or more, compared to
the tree contribution, and should be taken into account
in LHC computations.

These features make the neutralino pair  production processes
rather  interesting for testing the SUSY
dynamics at LHC. The reason is that they provide
tests which will be complementary to those addressing the cascade decays
of initially produced colored
SUSY particles to eventually $\tchi^0_1$, which is here assumed  to be the
LSP;  \eg studies of  mass spectra and decay branching
ratios \cite{SUSYsearches2}.  In particular,  consistency checks
should thus become available, allowing the strengthening
of possible constraints on  the validity of specific models.

Moreover, in such neutralino pair production, the role of the
 Majorana nature of the final state particles is more
  prominent than in decays involving just one neutralino at a time.
Since no  such states (except possible the neutrinos)
have been observed in the past, it would be interesting to have
eventually some experimental support of
our understanding of the Majorana nature.

If $\tchi^0_1$ turns out to be  an important or dominant component
Dark Matter, the present calculations\footnote{For the same reason,
Linear Collider studies should also be helpful \cite{gammagammaDM}.},
should  also help in providing LHC   constraints
on the direct or indirect observations of the Dark Matter properties,
whenever they will become available.

\appendix
\newpage

\noindent
{\Large \bf Appendix A}

\renewcommand{\theequation}{A.\arabic{equation}}
\renewcommand{\thesection}{A.\arabic{section}}
\setcounter{equation}{0}
\setcounter{section}{0}

\section{Tree level  helicity amplitudes for $q\bar q \to \tchi_i^0\tchi_j^0$}

\begin{figure}[t]
\vspace*{-3cm}
\[
\hspace{-1.cm}\epsfig{file=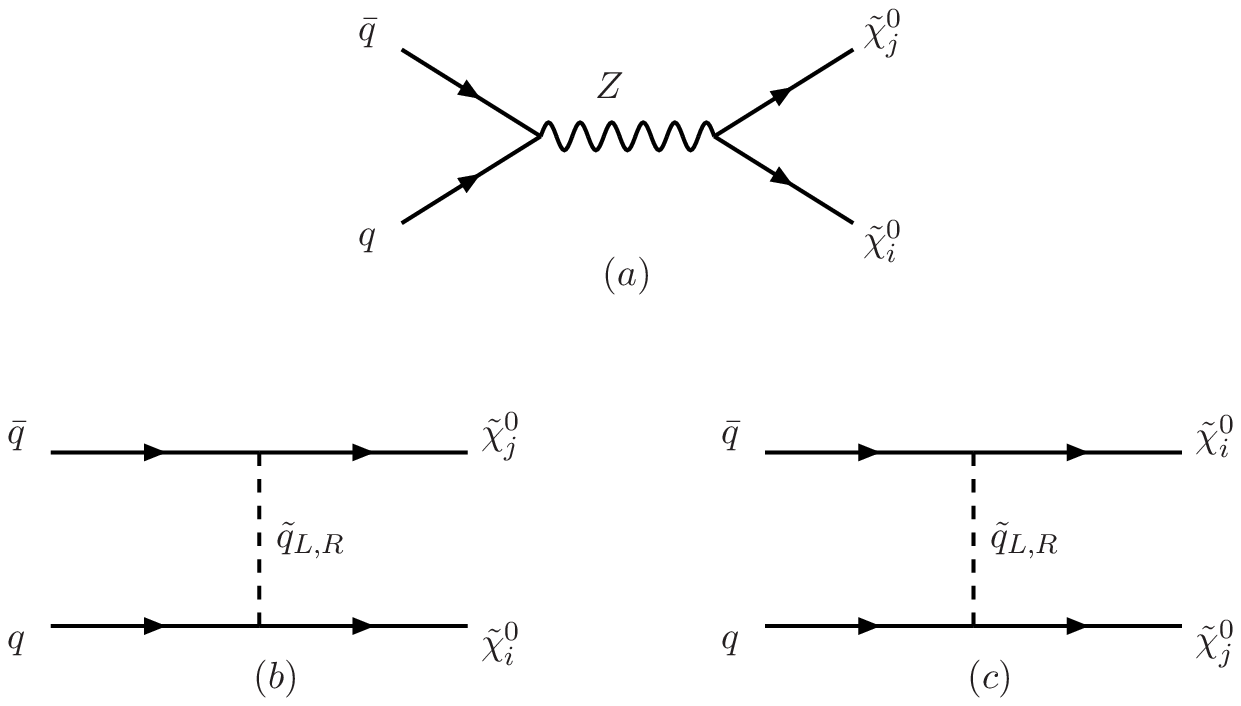,height=8.cm, width=13.cm}
\]
\vspace*{-1cm}
\caption[1]{ Feynman diagrams for $q\bar q\to\tchi^0_i\tchi^0_j$.}
\label{qq-diagrams}
\end{figure}

\begin{figure}[b]
\vspace*{-2cm}
\[
\hspace{-1.cm}\epsfig{file=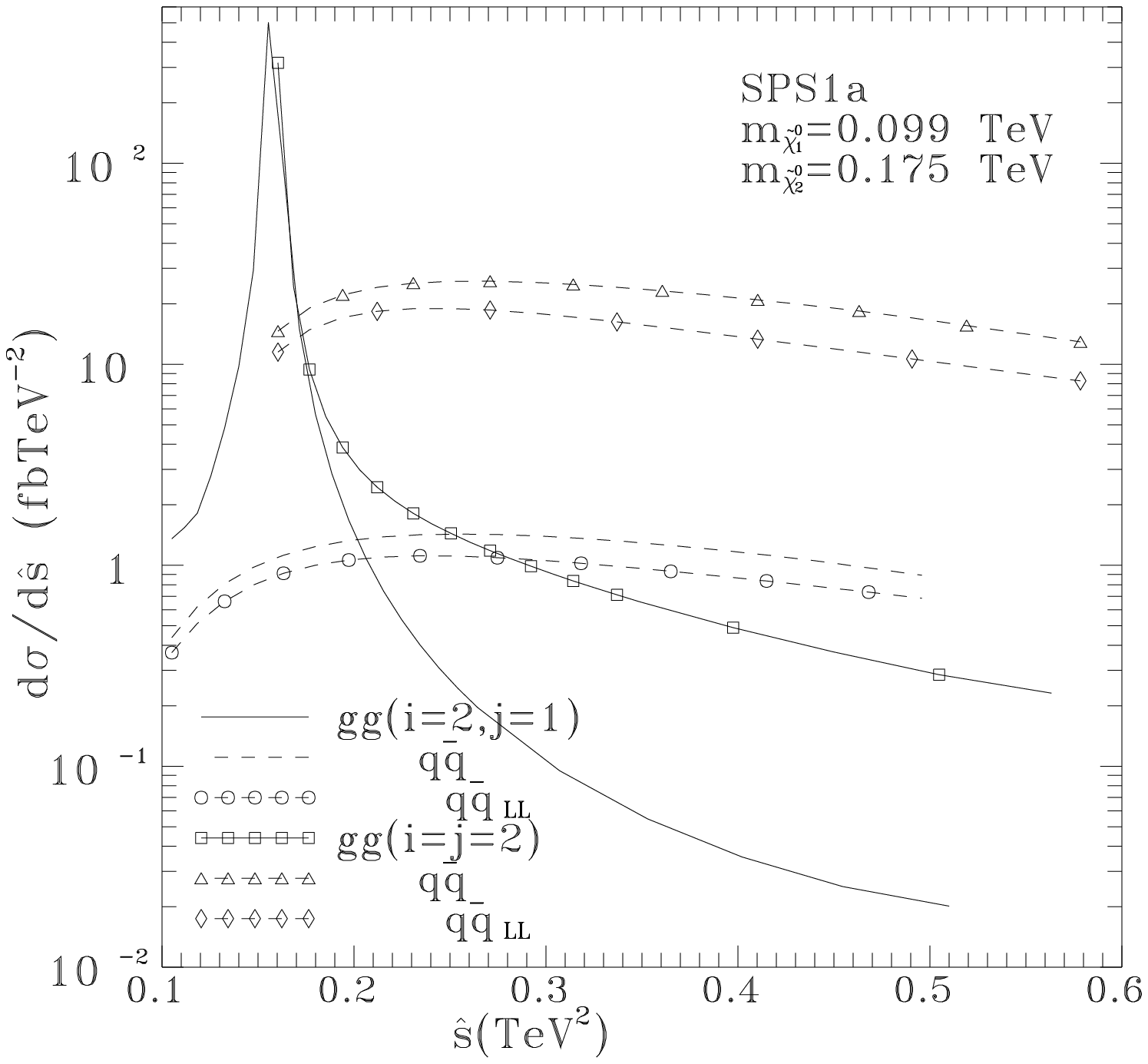,height=9.cm}
\]
\vspace*{-1cm}
\caption[1]{The $\shat$ distributions in the SPS1a model of \cite{Snowmass}.}
\label{SPS1a-fig}
\end{figure}

\begin{figure}[t]
\vspace*{-2cm}
\[
\hspace{-1.cm}\epsfig{file=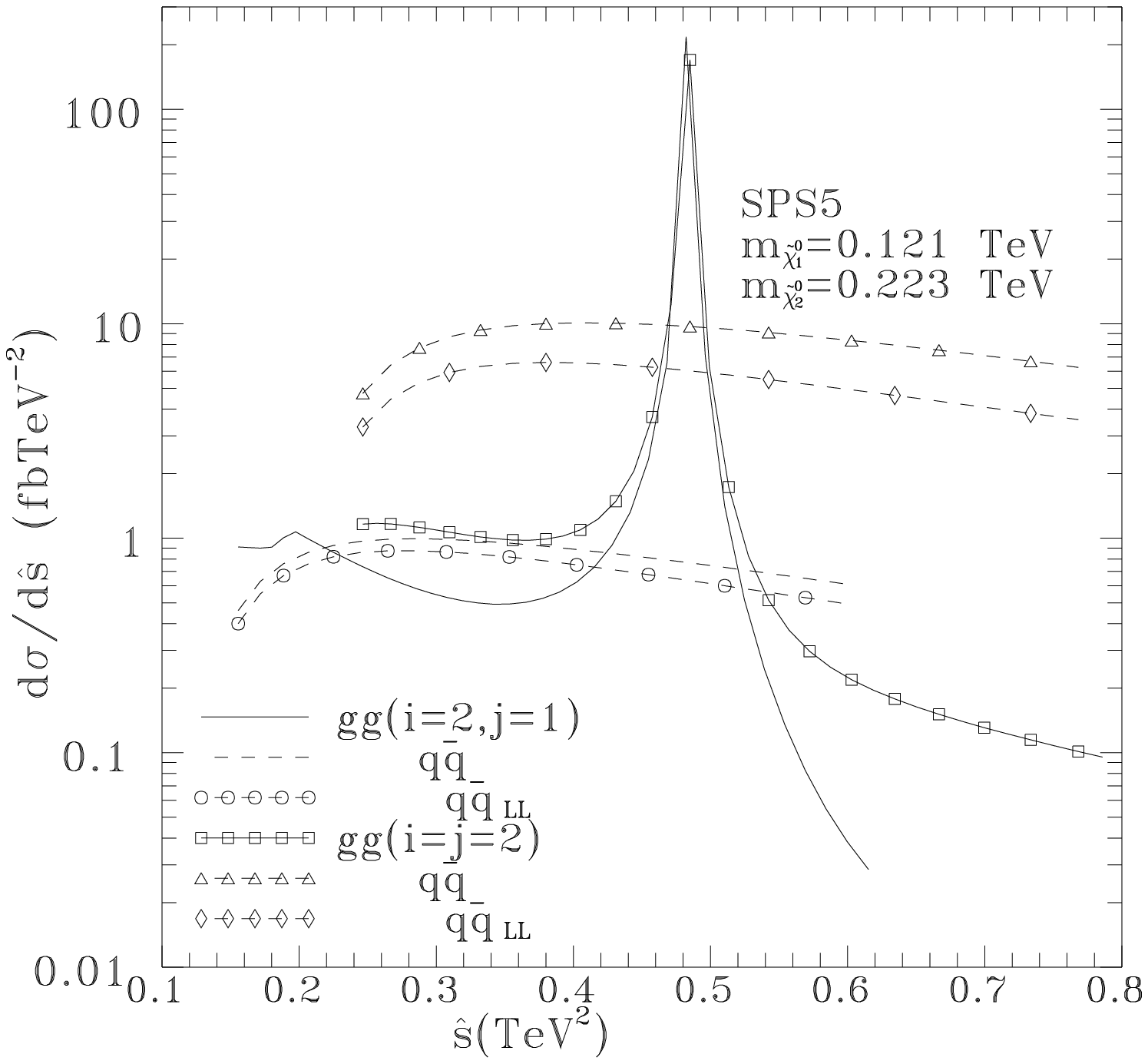,height=9.cm}
\]
\vspace*{-1cm}
\caption[1]{The $\shat$ distributions in the SPS5 model of \cite{Snowmass}. }
\label{SPS5-fig}
\end{figure}

\begin{figure}[b]
\vspace*{-1cm}
\[
\hspace{-0.5cm}\epsfig{file=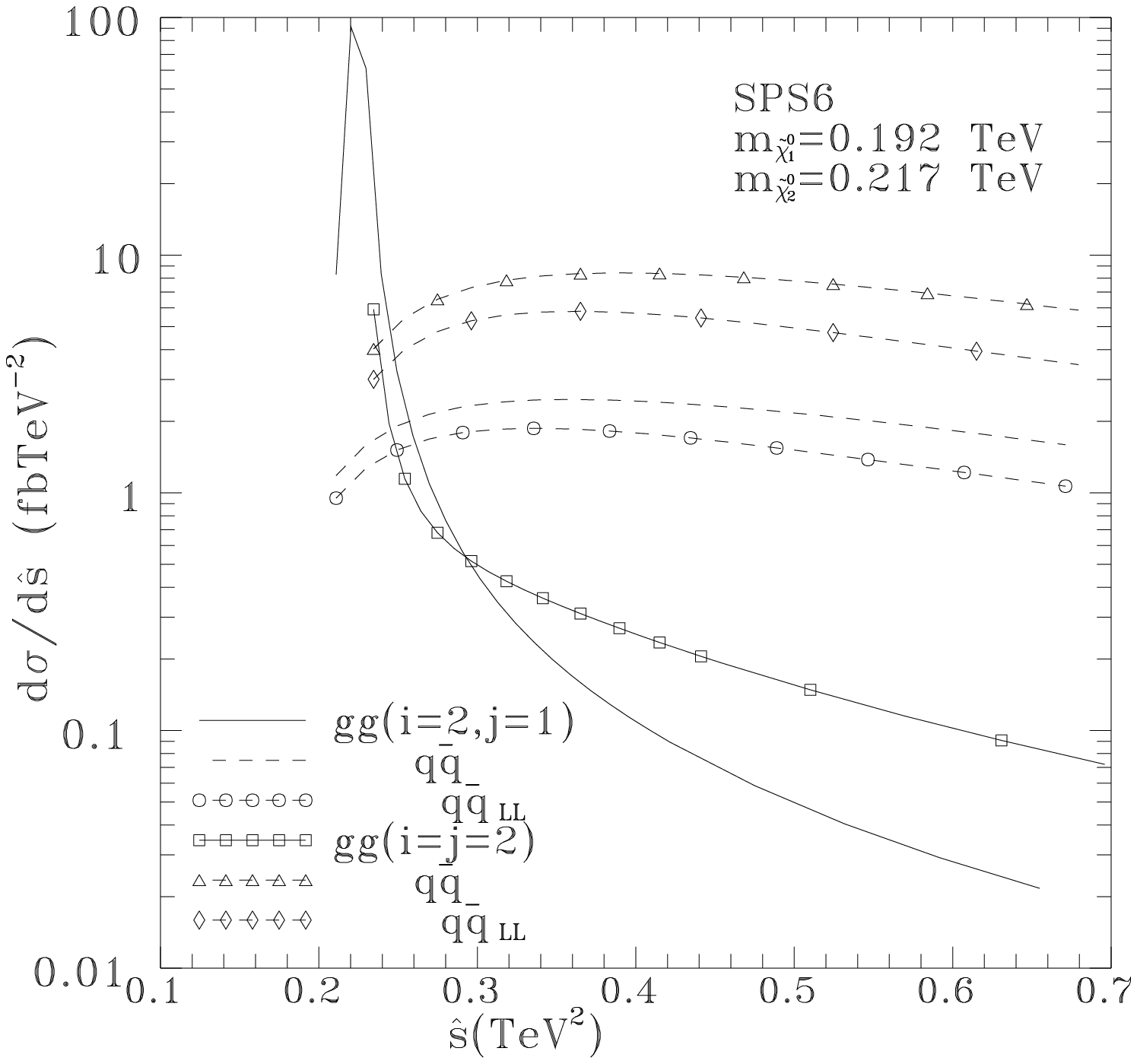,height=9.cm}
\]
\vspace*{-1cm}
\caption[1]{The $\shat$ distributions in the SPS6 model of \cite{Snowmass}.}
\label{SPS6-fig}
\end{figure}

\begin{figure}[t]
\vspace*{-2cm}
\[
\hspace{-1.cm}\epsfig{file=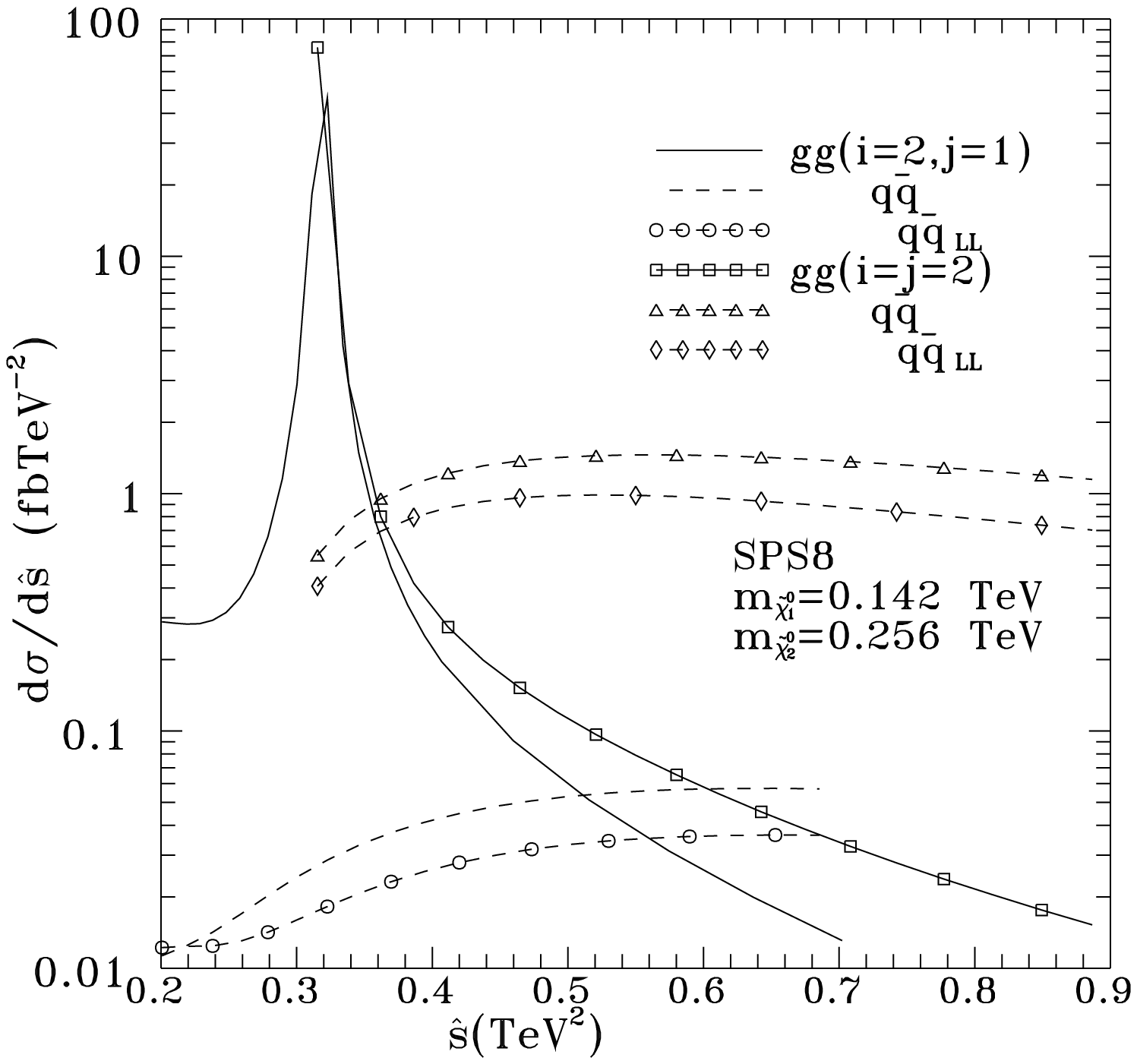,height=9.cm}
\]
\vspace*{-1cm}
\caption[1]{The $\shat$ distributions in the SPS8 model of \cite{Snowmass}.}
\label{SPS8-fig}
\end{figure}

\begin{figure}[b]
\vspace*{-1cm}
\[
\hspace{-1.cm}\epsfig{file=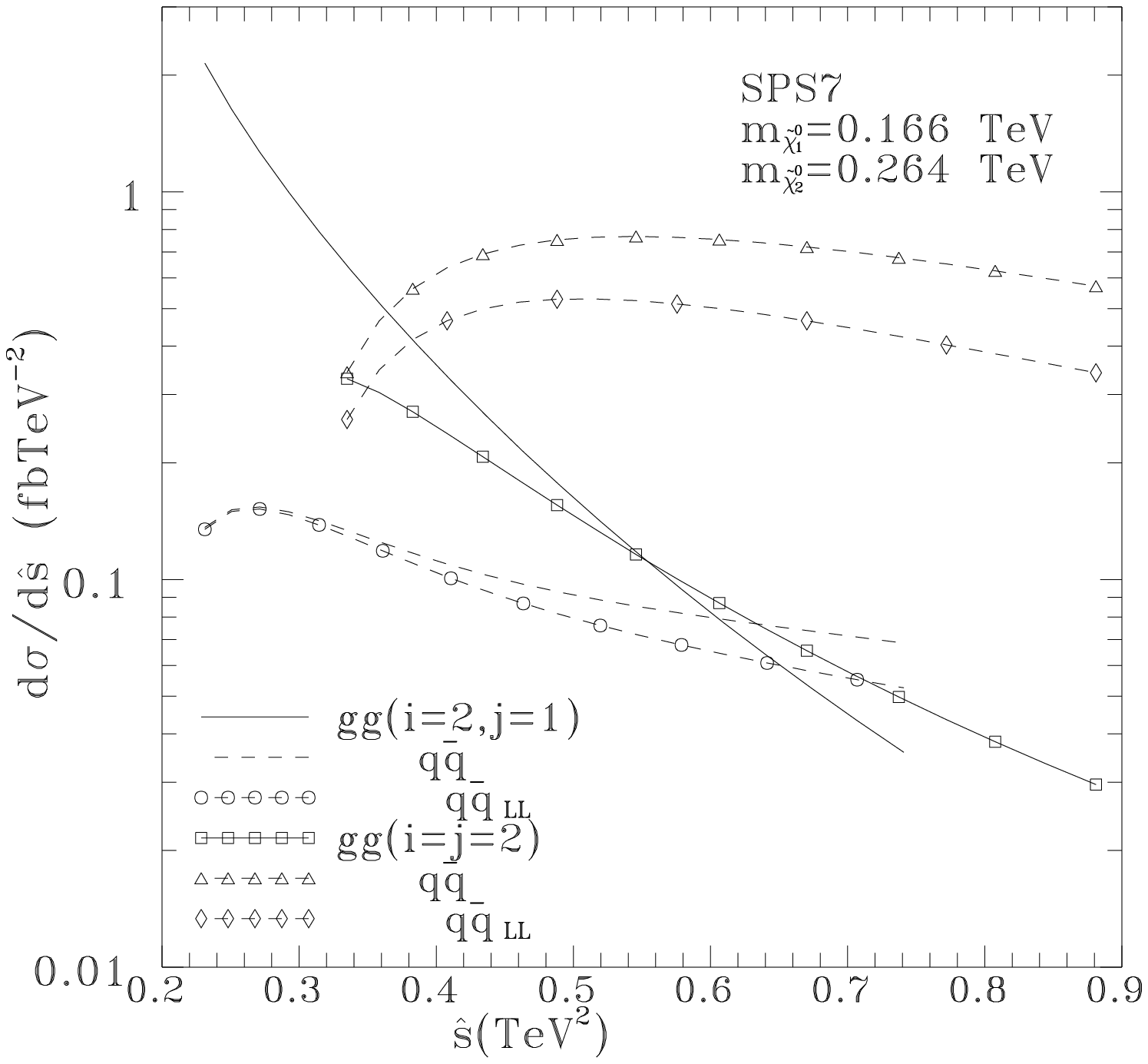,height=9.cm}
\]
\vspace*{-1cm}
\caption[1]{The $\shat$ distributions in the SPS7 model of \cite{Snowmass}.}
\label{SPS7-fig}
\end{figure}

\begin{figure}[t]
\vspace*{-2cm}
\[
\hspace{-1.cm}\epsfig{file=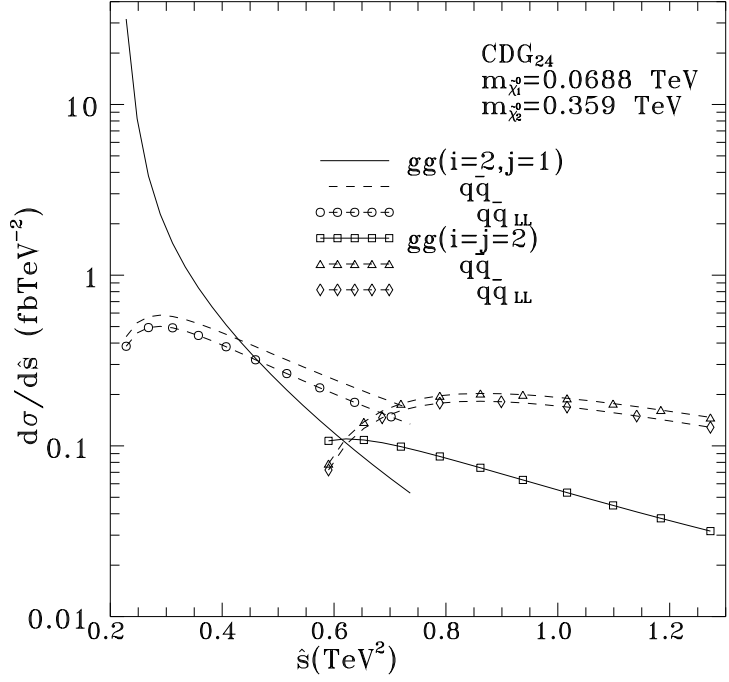,height=9.cm}
\]
\vspace*{-1cm}
\caption[1]{The $\shat$ distributions in the $CDG_{24}$ model of \cite{CDG}. }
\label{CDG24-fig}
\end{figure}

\begin{figure}[b]
\vspace*{-1cm}
\[
\hspace{-1.cm}\epsfig{file=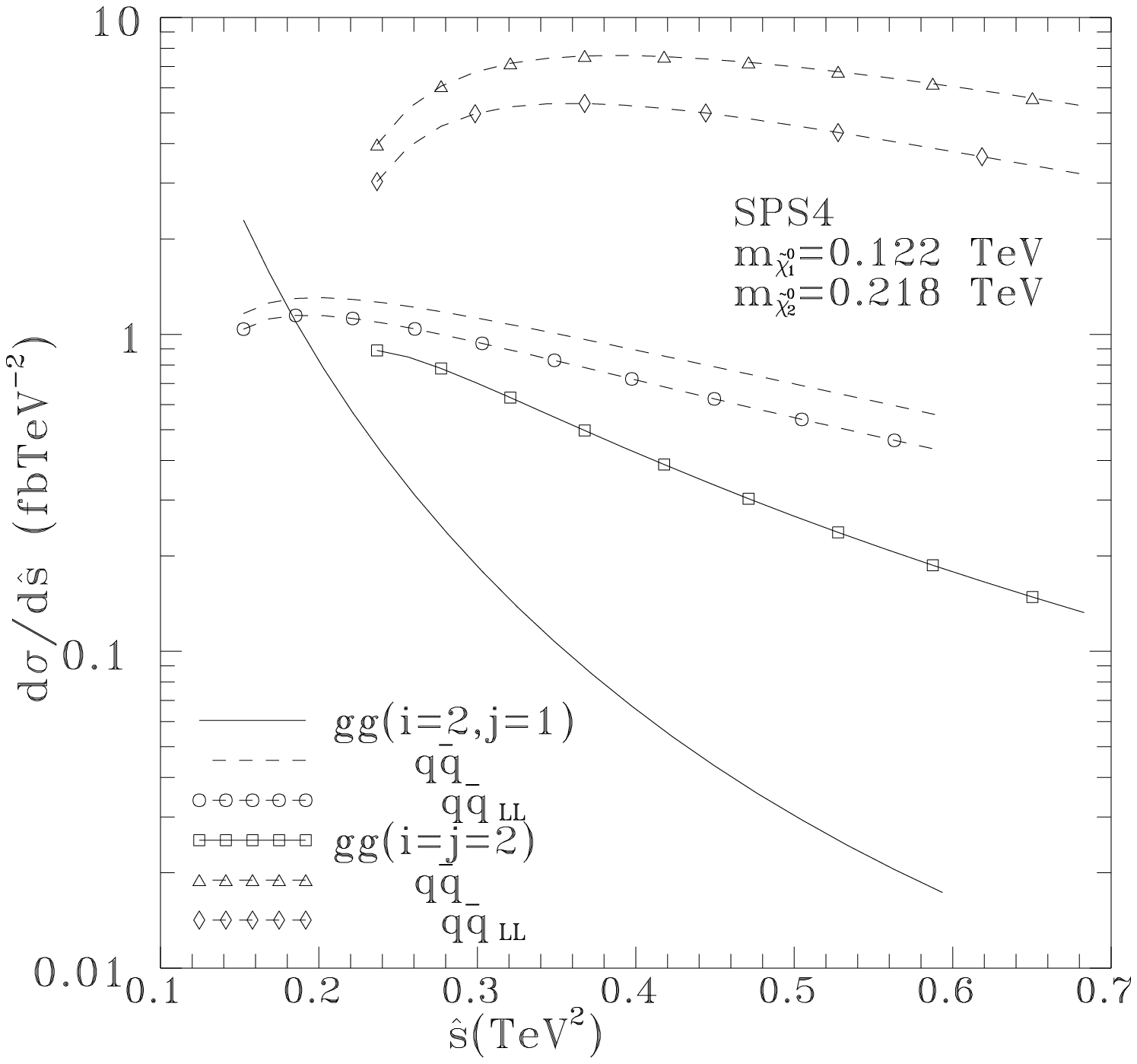,height=9.cm}
\]
\vspace*{-1cm}
\caption[1]{The $\shat$ distributions in the SPS4 model of \cite{Snowmass}.}
\label{SPS4-fig}
\end{figure}

Using the notation of (\ref{born}), the
  Born contributions arising from  the s-channel diagram in Fig.\ref{qq-diagrams}a,
  to the  seven basic helicity amplitudes   listed in (\ref{basic7})
 consist of
\bqa
F^{ijBs}_{-+-+}&=&{e^2g_{qL}\sqrt{\shat} \over 8s^2_Wc^2_W(\shat -m^2_Z)}
(1+\cos\theta^*)\Big[\sqrt{\shat -(m_i-m_j)^2}(N^{ij}-N^{ij*})
\nonumber \\
&& +\sqrt{\shat -(m_i+m_j)^2}(N^{ij}+N^{ij*})\Big ] \nonumber ~~, \\
F^{ijBs}_{+-+-}&=&{e^2g_{qR}\sqrt{\shat} \over 8s^2_Wc^2_W(\shat -m^2_Z)}
(1+\cos\theta^*)\Big[\sqrt{\shat -(m_i-m_j)^2}(N^{ij}-N^{ij*})
\nonumber \\
&& -\sqrt{\shat -(m_i+m_j)^2}(N^{ij}+N^{ij*}) \Big ] \nonumber ~~, \\
F^{ijBs}_{-+++}&=& {e^2g_{qL} \over 8s^2_Wc^2_W(\shat -m^2_Z)}
\sin\theta^* \Big[\sqrt{\shat -(m_i-m_j)^2}(m_i+m_j)(N^{ij}-N^{ij*})
\nonumber \\
&& +\sqrt{\shat -(m_i+m_j)^2}(m_i-m_j)(N^{ij}+N^{ij*})\Big ] \nonumber ~~, \\
F^{ijBs}_{+-++}&=&- {e^2g_{qR} \over 8s^2_Wc^2_W(\shat -m^2_Z)}
\sin\theta^* \Big[\sqrt{\shat -(m_i-m_j)^2}(m_i+m_j)(N^{ij}-N^{ij*})
\nonumber \\
&& +\sqrt{\shat -(m_i+m_j)^2}(m_i-m_j)(N^{ij}+N^{ij*})\Big ]    ~~,
\label{Born-s-basic1}
\eqa
while
\bq
F^{ijBs}_{++\tau_i\tau_j}=F^{ijBs}_{--\tau_i\tau_j}=0~~, \label{Born-s-basic2}
\eq
for all $(\tau_i,~\tau_j)$-values\footnote{In fact, the
  s-channel Born contributions vanish
for all amplitudes  with equal incoming helicities.}.
Here   $\theta^*$  is the   scattering angle in the c.m. of the subprocess,
and   $Z^N$  denotes
 the neutralino mixing matrix  in the notation
of \cite{Rosiek}.

The   Born contributions to the  seven basic helicity amplitudes
 of  (\ref{basic7}),   arising from  the t-channel diagram
 in Fig.\ref{qq-diagrams}b, are
\bqa
F^{ijBt}_{-+-+}&=&-\frac{\sqrt{\shat}}{4}
\Big [\sqrt{\shat-(m_i-m_j)^2}-\sqrt{\shat-(m_i+m_j)^2}\Big ]
\nonumber \\
&& \cdot \sum_n \left ({1\over \hat t-\tilde m^2_n}
A^{L}_i(\tilde q_n)A^{L*}_j(\tilde q_n)\right )
(1+\cos \theta^*)~, \nonumber \\
F^{ijBt}_{+-+-}&=&-\frac{\sqrt{\shat}}{4}
\Big [\sqrt{\shat-(m_i-m_j)^2}-\sqrt{\shat-(m_i+m_j)^2}\Big ]
\nonumber \\
&& \cdot \sum_n \left ({1\over \hat t-\tilde m^2_n}
A^{R}_i(\tilde q_n)A^{R*}_j(\tilde q_n)\right )
(1+\cos \theta^*)~, \nonumber \\
F^{ijBt}_{-+++}&=&-\frac{1}{4}
\Big [(m_i+m_j) \sqrt{\shat-(m_i-m_j)^2}-(m_i-m_j)\sqrt{\shat-(m_i+m_j)^2}\Big ]
\nonumber \\
&& \cdot \sum_n \left ({1\over \hat t-\tilde m^2_n}
A^{L}_i(\tilde q_n)A^{L*}_j(\tilde q_n)\right )
\sin \theta^* ~, \nonumber \\
F^{ijBt}_{+-++}&=& \frac{1}{4}
\Big [(m_i+m_j) \sqrt{\shat-(m_i-m_j)^2}+(m_i-m_j)\sqrt{\shat-(m_i+m_j)^2}\Big ]
\nonumber \\
&& \cdot \sum_n \left ({1\over \hat t-\tilde m^2_n}
A^{R}_i(\tilde q_n)A^{R*}_j(\tilde q_n) \right )
\sin \theta^* ~, \nonumber \\
F^{ijBt}_{++++}&=&-\frac{\sqrt{\shat}}{4}
\Big [\sqrt{\shat-(m_i-m_j)^2}-\sqrt{\shat-(m_i+m_j)^2}\Big ]
\nonumber \\
&& \cdot \sum_n \left ({1\over \hat t-\tilde m^2_n}
A^{R}_i(\tilde q_n)A^{L*}_j(\tilde q_n)) \right )
(1+\cos \theta^* )~, \nonumber \\
F^{ijBt}_{++--}&=& \frac{\sqrt{\shat}}{4}
\Big [\sqrt{\shat-(m_i-m_j)^2}+\sqrt{\shat-(m_i+m_j)^2}\Big ]
\nonumber \\
&& \cdot \sum_n \left ({1\over \hat t-\tilde m^2_n}
A^{R}_i(\tilde q_n)A^{L*}_j(\tilde q_n)\right )
(1-\cos \theta^* )~, \nonumber \\
F^{ijBt}_{+++-}&=&-\frac{1}{4}
\Big [(m_i+m_j) \sqrt{\shat-(m_i-m_j)^2}+(m_i-m_j)\sqrt{\shat-(m_i+m_j)^2}\Big ]
\nonumber \\
&& \cdot \sum_n \left ({1\over \hat t-\tilde m^2_n}
A^{R}_i(\tilde q_n)A^{L*}_j(\tilde q_n)\right )
\sin \theta^* ~, \label{Born-t-basic}
\eqa
while  the corresponding contributions by
the u-channel diagram in Fig.\ref{qq-diagrams}c,
are
\bqa
F^{ijBu}_{-+-+}&=& \frac{\sqrt{\shat}}{4}
\Big [\sqrt{\shat-(m_i-m_j)^2} +\sqrt{\shat-(m_i+m_j)^2}\Big ]
\nonumber \\
&& \cdot \sum_n \left ({1\over \uhat-\tilde m^2_n}
A^{L}_j(\tilde q_n)A^{L*}_i(\tilde q_n)\right )
(1+\cos \theta^* )~, \nonumber \\
F^{ijBu}_{+-+-}&=& \frac{\sqrt{\shat}}{4}
\Big [\sqrt{\shat-(m_i-m_j)^2}+\sqrt{\shat-(m_i+m_j)^2}\Big ]
\nonumber \\
&& \cdot \sum_n \left ({1\over \uhat-\tilde m^2_n}
A^{R}_j(\tilde q_n)A^{R*}_i(\tilde q_n)\right )
(1+\cos \theta^* )~, \nonumber \\
F^{ijBu}_{-+++}&=& \frac{1}{4}
\Big [(m_i+m_j) \sqrt{\shat-(m_i-m_j)^2}+(m_i-m_j)\sqrt{\shat-(m_i+m_j)^2}\Big ]
\nonumber \\
&& \cdot \sum_n \left ({1\over \uhat-\tilde m^2_n}
A^{L}_j(\tilde q_n)A^{L*}_i(\tilde q_n) \right )
\sin \theta^* ~, \nonumber \\
F^{ijBu}_{+-++}&=& -\frac{1}{4}
\Big [(m_i+m_j) \sqrt{\shat-(m_i-m_j)^2}-(m_i-m_j)\sqrt{\shat-(m_i+m_j)^2}\Big ]
\nonumber \\
&& \cdot \sum_n \left ({1\over \uhat-\tilde m^2_n}
A^{R}_j(\tilde q_n)A^{R*}_i(\tilde q_n) \right )
\sin \theta^* ~, \nonumber \\
F^{ijBu}_{++++}&=&-\frac{\sqrt{\shat}}{4}
\Big [\sqrt{\shat-(m_i-m_j)^2}-\sqrt{\shat-(m_i+m_j)^2}\Big ]
\nonumber \\
&& \cdot \sum_n \left ({1\over \uhat-\tilde m^2_n}
A^{R}_j(\tilde q_n)A^{L*}_i(\tilde q_n)\right )
(1-\cos \theta^* )~, \nonumber \\
F^{ijBu}_{++--}&=& \frac{\sqrt{\shat}}{4}
\Big [\sqrt{\shat-(m_i-m_j)^2}+\sqrt{\shat-(m_i+m_j)^2}\Big ]
\nonumber \\
&& \cdot \sum_n \left ({1\over \uhat-\tilde m^2_n}
A^{R}_j(\tilde q_n)A^{L*}_i(\tilde q_n)\right )
(1+\cos \theta^* )~, \nonumber \\
F^{ijBu}_{+++-}&=& \frac{1}{4}
\Big [(m_i+m_j) \sqrt{\shat-(m_i-m_j)^2}+(m_i-m_j)\sqrt{\shat-(m_i+m_j)^2}\Big ]
\nonumber \\
&& \cdot \sum_n \left ({1\over \uhat-\tilde m^2_n}
A^{R}_j(\tilde q_n)A^{L*}_i(\tilde q_n)\right )
\sin \theta^* ~ \label{Born-u-basic}.
\eqa
The summation in (\ref{Born-t-basic}, \ref{Born-u-basic})
runs over the left and right squarks with the same flavor as the
incoming quarks and  mass $\tilde m_n$.
The couplings in (\ref{Born-s-basic1}, \ref{Born-t-basic},
\ref{Born-u-basic}) have been defined in (\ref{Zqq1}-\ref{qqchi2}).

\section{Leading log  helicity amplitudes for $q\bar q \to \tchi_i^0\tchi_j^0$.}

In this subsection we include the one-loop leading log contributions to
the seven basic amplitudes  of (\ref{basic7}).
We use the  various couplings  defined
  in (\ref{Zqq1}-\ref{qqchi2}) and
\bqa
&&\bar m_d\equiv \frac{m_d}{\cos\beta} ~~,~~
 \bar m_u\equiv \frac{m_u}{\sin\beta} ~~, \label{Yukawa-apB}  \\
&& Z^N_{di}\equiv Z^N_{3i}~~,~~ Z^N_{ui}\equiv Z^N_{4i}~~,~~
g=e/s_W,~g'=e/c_W~~, \label{chi0-mixing-apB} \\
&& \tilde \beta_0= \frac{3}{4} C_A- \frac{n_g}{2}-\frac{n_h}{8}
=-~{1\over4}  ~~~~,~~~ \tilde \beta_0^\prime =-\frac{5}{6}n_g-\frac{n_h}{8}
=-~{11\over4}~~~,  \label{beta-apB}
\eqa
while   an $\ln$-symbol standing alone should be understood as
\bq
 \ln \to \ln\frac{\shat}{\mwd} ~~.
\eq

Denoting then  by  $(q=u,~d)$  the quark
occurring in the initial state, and
by $(q'=d,~u)$ the corresponding companion quark belonging to
the same SU(2) doublet, and describing by $M_S$ the  effective average mass for the
the squarks of  the same flavor as the incoming quarks, the
  seven basic amplitudes of (\ref{basic7})  are written as
\bqa
&& F^{ij}_{-+-+}(\theta^*)=F^{ijB}_{-+-+}(\theta^*)
\Bigg [{\alpha(1+26c^2_W)\over144\pi s^2_Wc^2_W}(2\ln-\ln^2)
-~{\alpha_s\over3\pi}\ln{\hat s\over \mw^2}\nonumber\\
&&
-~{\alpha[\ln]\over8\pi s^2_W \mwd } \Big ({m^2_t\over  \sin^2\beta }
+{m^2_b\over \cos^2\beta} \Big )(\delta_{qt}+\delta_{qb})\Bigg ]
  +F^{ijBs}_{-+-+}(\theta^*)
 \Big[{\alpha(1+2c^2_W)\over16\pi s^2_Wc^2_W}(2\ln-\ln^2)\Big ]
\nonumber\\
&&  -{3\alpha^2 g_{qL}\sqrt{\hat s}(1+\cos\theta^*)\over
16 s^4_Wc^2_W\mwd (\hat s-m^2_Z)}
[\ln]\Bigg \{  \sqrt{\hat s-(m_i-m_j)^2}
\Big [(Z^{N*}_{4i}Z^{N}_{4j}-Z^{N}_{4i}Z^{N*}_{4j}){m^2_t\over \sin^2\beta }
\nonumber\\
&& -(Z^{N*}_{3i}Z^{N}_{3j}-Z^{N}_{3i}Z^{N*}_{3j}){m^2_b\over \cos^2\beta }\Big ]
\nonumber\\
&&+
\sqrt{\hat s-(m_i+m_j)^2}
\Big [(Z^{N*}_{4i}Z^{N}_{4j}+Z^{N}_{4i}Z^{N*}_{4j}){m^2_t\over \sin^2\beta }
-(Z^{N*}_{3i}Z^{N}_{3j}+Z^{N}_{3i}Z^{N*}_{3j}){m^2_b\over \cos^2\beta } \Big ]
\Bigg \}
\nonumber\\
&&+I^{(3)}_{q}{\alpha^2[\ln^2]\sqrt{\hat s}\over 12 s^4_Wc_W}\Big \{~
{(1+\cos\theta^*)\over\hat t-\tilde m^2(\tilde q_L)}\Big [\sqrt{\hat s-(m_i-m_j)^2}-
\sqrt{\hat s-(m_i+m_j)^2}\Big ]\nonumber\\
&& \cdot [Z^{N*}_{2j}(Z^{N}_{1i}s_W+6I^{(3)}_{q}Z^{N}_{2i}c_W)
+Z^{N}_{2i}(Z^{N*}_{1j}s_W+6I^{(3)}_{q}Z^{N*}_{2j}c_W)]
\nonumber\\
&&
-~{(1+\cos\theta^*)\over\hat u-\tilde m^2(\tilde q_L)}\Big [\sqrt{\hat s-(m_i-m_j)^2}+
\sqrt{\hat s-(m_i+m_j)^2} \Big ]\nonumber\\
&& \cdot [Z^{N*}_{2i}(Z^{N}_{1j}s_W+6 I^{(3)}_{q}Z^{N}_{2j}c_W)
+Z^{N}_{2j}(Z^{N*}_{1i}s_W+ 6 I^{(3)}_{q}Z^{N*}_{2i}c_W)]\Big \}\nonumber\\
&&
-I^{(3)}_{q}{\alpha^2[\ln]\over \sqrt{\hat s} s^4_W}(1+\cos\theta^*)
\Bigg \{[\sqrt{\hat s-(m_i-m_j)^2}+\sqrt{\hat s-(m_i+m_j)^2}]\nonumber\\
&&
\cdot \Big ((2Z^{N*}_{2i}Z^{N}_{2j}+Z^{N*}_{q'i}Z^{N}_{q'j} )
\Big [\ln{-\hat t\over\hat s}\Big ]-
(2Z^{N*}_{2i}Z^{N}_{2j}+Z^{N*}_{qi}Z^{N}_{qj})
\Big [\ln{-\hat u\over\hat s}\Big ]\Big )
\nonumber\\
&&+\Big [\sqrt{\hat s-(m_i-m_j)^2}-\sqrt{\hat s-(m_i+m_j)^2}\Big ]\nonumber\\
&&
\cdot \Big ((2Z^{N}_{2i}Z^{N*}_{2j}+Z^{N}_{qi}Z^{N*}_{qj})
\Big [\ln{-\hat t\over\hat s}\Big ]
-(2Z^{N}_{2i}Z^{N*}_{2j}+Z^{N}_{q'i}Z^{N*}_{q'j}  )
\Big [\ln{-\hat u\over\hat s}\Big ]\Big )\Bigg \}
\nonumber\\
&&+ 2I^{(3)}_{q}\sqrt{\hat s}[\ln]\Bigg \{{(1+\cos\theta^*)
\over \hat t-M^2_S}
[\sqrt{\hat s-(m_i-m_j)^2}-\sqrt{\hat s-(m_i+m_j)^2}]\nonumber\\
&& \cdot \Bigg (
\Big [\ln{-\hat t\over \hat s}\Big ]\Big ( {\alpha^2\over12s^4_Wc_W}
 [ Z^{N}_{2i}(Z^{N*}_{1j}s_W+6 I^{(3)}_{q}Z^{N*}_{2j}c_W)
+Z^{N*}_{2j}(Z^{N}_{1i}s_W+ 6I^{(3)}_{q}Z^{N}_{2i}c_W) ]\nonumber\\
&&
~+~{\alpha^2{\bar m_{q}}^2\over4m^2_Ws^4_W}
Z^{N*}_{qj}Z^{N}_{qi} \Big ) \nonumber\\
&&
+\Big [\ln{-\hat u\over\hat s}\Big ] \Big ({\alpha^2\over12s^4_Wc_W}
[Z^{N*}_{2j}(Z^{N}_{1i}s_W- 6I^{(3)}_{q} Z^{N}_{2i}c_W)
+Z^{N}_{2i}(Z^{N*}_{1j}s_W-6I^{(3)}_{q}Z^{N*}_{2j}c_W )]\nonumber\\
&&
~-~{\alpha^2{\bar m_{q'}}^2\over4m^2_Ws^4_W}
Z^{N}_{q'i}Z^{N*}_{q'j}\Big ) \Bigg ) \nonumber\\
&&-~{(1+\cos\theta^*)\over \hat u -M^2_S}
[\sqrt{\hat s-(m_i-m_j)^2}+\sqrt{\hat s-(m_i+m_j)^2}]\nonumber\\
&&\Bigg ( \Big [\ln{-\hat u\over \hat s}\Big ] \Big ({\alpha^2\over12s^4_Wc_W}
[ Z^{N}_{2j}(Z^{N*}_{1i}s_W+ 6 I^{(3)}_{q}Z^{N*}_{2i}c_W)
+Z^{N*}_{2i}(Z^{N}_{1j}s_W+6 I^{(3)}_{q}Z^{N}_{2j}c_W)]\nonumber\\
&&
~+~{\alpha^2{\bar m_{q}}^2\over4m^2_Ws^4_W}
Z^{N*}_{qi}Z^{N}_{qj} \Big )  \nonumber\\
&&
+\Big [\ln{-\hat t\over\hat s}\Big ]\Big ({\alpha^2\over12s^4_Wc_W}
[Z^{N*}_{2i}(Z^{N}_{1j}s_W-6I^{(3)}_{q}Z^{N}_{2j}c_W)
+Z^{N}_{2j}(Z^{N*}_{1i}s_W-6I^{(3)}_{q}Z^{N*}_{2i}c_W)]\nonumber\\
&&
~-~{\alpha^2{\bar m_{q'}}^2\over4m^2_Ws^4_W}
Z^{N}_{q'j}Z^{N*}_{q'i}\Big )\Bigg ) \Bigg \}\nonumber\\
&&
-\Big [\frac{\tilde \beta_0}{\sw^4}
+\frac{\tilde \beta'_0}{\cw^4}(1-2|Q_q|)\Big ]
{I^{(3)}_{q}\alpha^2 \sqrt{\hat s}\over (\hat s-m^2_Z)}(1+\cos\theta^*)
[(N^{ij}-N^{ij*})\sqrt{\hat s-(m_i-m_j)^2}
\nonumber \\
&& + (N^{ij}+N^{ij*})\sqrt{\hat s-(m_i+m_j)^2})][\ln]~~, \label{LL-+-+}\\[0.5cm]
&& F^{ij}_{-+++}(\theta^*)=F^{ijB}_{-+++}(\theta^*)
\Big [{\alpha(1+26c^2_W)\over144\pi s^2_Wc^2_W}(2\ln-\ln^2)
-~{\alpha_s\over3\pi}\ln{\hat s\over \mw^2}\nonumber\\
&&
-~{\alpha[\ln]\over8\pi s^2_W \mwd}({m^2_t\over \sin^2\beta }
+{m^2_b\over \cos^2\beta })(\delta_{qt}+\delta_{qb})\Big ]
+F^{ijBs}_{-+++}(\theta^*)\Big [{\alpha(1+2c^2_W)\over16\pi s^2_Wc^2_W}
(2\ln-\ln^2)\Big ]
\nonumber\\
&&-{3 \alpha^2 g_{qL}\sin\theta^*\over 16 s^4_Wc^2_W \mwd (\hat s-m^2_Z)}
[\ln] \Big \{ (m_i+m_j)\sqrt{\hat s-(m_i-m_j)^2}
\Big [(Z^{N*}_{4i}Z^{N}_{4j}-Z^{N}_{4i}Z^{N*}_{4j}){m^2_t\over \sin^2\beta }
\nonumber\\
&&-(Z^{N*}_{3i}Z^{N}_{3j}-Z^{N}_{3i}Z^{N*}_{3j}){m^2_b\over \cos^2\beta }\Big ]
- (m_i-m_j)\sqrt{\hat s-(m_i+m_j)^2}
\Big [(Z^{N*}_{4i}Z^{N}_{4j}+Z^{N}_{4i}Z^{N*}_{4j}){m^2_t\over \sin^2\beta }
\nonumber\\
&&-(Z^{N*}_{3i}Z^{N}_{3j}+Z^{N}_{3i}Z^{N*}_{3j}){m^2_b\over \cos^2\beta} \Big ]\Big \}
\nonumber\\
&&+I^{(3)}_{q}{\alpha^2 [\ln^2]\over 12  s^4_Wc_W}
\Big \{ {\sin\theta^*\over\hat t-\tilde m^2(\tilde q_L)}
[(m_i+m_j)\sqrt{\hat s-(m_i-m_j)^2}-
(m_i-m_j)\sqrt{\hat s-(m_i+m_j)^2}]
\nonumber\\
&&\cdot [Z^{N*}_{2j}(Z^{N}_{1i}s_W+ 6 I^{(3)}_{q}Z^{N}_{2i}c_W)
+Z^{N}_{2i}(Z^{N*}_{1j}s_W+ 6 I^{(3)}_{q}Z^{N*}_{2j}c_W)]
\nonumber\\
&& -{\sin\theta^*\over\hat u-\tilde m^2(\tilde q_L)}
[(m_i+m_j)\sqrt{\hat s-(m_i-m_j)^2}+
(m_i-m_j)\sqrt{\hat s-(m_i+m_j)^2}]\nonumber\\
&& \cdot [Z^{N*}_{2i}(Z^{N}_{1j}s_W+3(2I^{(3)}_{q})Z^{N}_{2j}c_W)
+Z^{N}_{2j}(Z^{N*}_{1i}s_W+3(2I^{(3)}_{q})Z^{N*}_{2i}c_W)] \Big \}
\nonumber\\
&& -I^{(3)}_{q} {\alpha^2[\ln]\over \hat s s^4_W}\sin\theta^*
\Big \{[(m_i+m_j)\sqrt{\hat s-(m_i-m_j)^2}
+(m_i-m_j)\sqrt{\hat s-(m_i+m_j)^2}]\nonumber\\
&&
\cdot \Big ((2Z^{N*}_{2i}Z^{N}_{2j}+Z^{N*}_{q'i}Z^{N}_{q'j})
\Big [\ln{-\hat t\over\hat s}\Big ]-
(2Z^{N*}_{2i}Z^{N}_{2j}+Z^{N*}_{qi}Z^{N}_{qj})\Big [\ln{-\hat u\over\hat s}\Big ]\Big )
\nonumber\\
&&+[(m_i+m_j)\sqrt{\hat s-(m_i-m_j)^2}
-(m_i-m_j)\sqrt{\hat s-(m_i+m_j)^2}]
\nonumber\\
&&
\cdot \Big ((2Z^{N}_{2i}Z^{N*}_{2j}+Z^{N}_{qi}Z^{N*}_{qj})
\Big [\ln{-\hat t\over\hat s}\Big ]
-(2Z^{N}_{2i}Z^{N*}_{2j}+Z^{N}_{q'i}Z^{N*}_{q'j})
\Big [\ln{-\hat u\over\hat s}\Big ]\Big )\Big \}
\nonumber\\
&&+I^{(3)}_{q}{\alpha^2[\ln]\over 6 s^4_Wc_W}\Bigg \{{\sin\theta^*
\over \hat t-M^2_S}
[(m_i+m_j)\sqrt{\hat s-(m_i-m_j)^2}-(m_i-m_j)\sqrt{\hat s-(m_i+m_j)^2}]\nonumber\\
&& \cdot \Bigg ( \Big [\ln{-\hat t\over \hat s}\Big ]
\Big  [ Z^{N}_{2i}( Z^{N*}_{1j}s_W+6 I^{(3)}_{q}Z^{N*}_{2j}c_W)
+Z^{N*}_{2j}(Z^{N}_{1i}s_W+ 6 I^{(3)}_{q}Z^{N}_{2i}c_W)
 +{3{\bar m_{q}}^2c_W\over m^2_W}
Z^{N}_{qj}Z^{N*}_{qi} \Big ]\nonumber\\
&&
+\Big [\ln{-\hat u\over\hat s}\Big ]
\Big [ Z^{N*}_{2j}(Z^{N}_{1i}s_W-6 I^{(3)}_{q}Z^{N}_{2i}c_W)
+Z^{N}_{2i}(Z^{N*}_{1j}s_W-6 I^{(3)}_{q}Z^{N*}_{2j}c_W)
-{3 {\bar m_{q'}}^2\cw \over \mwd}Z^{N}_{q'i}Z^{N*}_{q'j} \Big ]\Bigg )
\nonumber\\
&&-{\sin\theta^*\over \hat u-M^2_S}
[(m_i+m_j)\sqrt{\hat s-(m_i-m_j)^2}+(m_i-m_j)\sqrt{\hat s-(m_i+m_j)^2}]
\nonumber\\
&& \cdot \Bigg ( \Big [\ln{-\hat u\over \hat s} \Big ]
\Big [ Z^{N}_{2j}(Z^{N*}_{1i}s_W+6 I^{(3)}_{q} Z^{N*}_{2i}c_W)
+Z^{N*}_{2i}(Z^{N}_{1j}s_W+ 6I^{(3)}_{q} Z^{N}_{2j}c_W )
+{3 {\bar m_{q}}^2\cw \over \mwd}Z^{N*}_{qi}Z^{N}_{qj}\Big ]
\nonumber\\
&&
+\Big [\ln{-\hat t\over\hat s}\Big ]
\Big [ Z^{N*}_{2i}(Z^{N}_{1j}s_W- 6 I^{(3)}_{q}Z^{N}_{2j}c_W)
+Z^{N}_{2j}(Z^{N*}_{1i}s_W- 6I^{(3)}_{q}Z^{N*}_{2i}c_W)
-{3 {\bar m_{q'}}^2\cw \over \mwd } Z^{N}_{q'j}Z^{N*}_{q'i}
\Big ] \bigg )\Bigg \}\nonumber\\
&&
-\Big [\frac{\tilde \beta_0}{\sw^4} +\frac{\tilde \beta'_0}{\cw^4} (1-2|Q_q|)\Big ]
{\alpha^2 I^{(3)}_{q}\over (\hat s-m^2_Z)}\sin\theta^*
[(N^{ij}-N^{ij*}) (m_i+m_j)\sqrt{\hat s-(m_i-m_j)^2}
\nonumber \\
&& +(N^{ij}+N^{ij*}) (m_i-m_j)\sqrt{\hat s-(m_i+m_j)^2}][\ln]
~~, \label{LL-+++}\\[0.5cm]
&& F^{ij}_{+-++}(\theta^*)=F^{ijB}_{+-++}(\theta^*)
\Big [{\alpha Q^2_q\over4\pi c^2_W}(2\ln-\ln^2)
-{\alpha[\ln]\over4\pi s^2_W\mwd }
\Big ({m^2_t\over \sin^2\beta }\delta_{qt}
+{m^2_b\over \cos^2\beta }\delta_{qb}\Big )
\nonumber\\
&& -{\alpha_s\over3\pi}\ln{\hat s\over \mw^2}\Big ]
+F^{ijBs}_{+-++}(\theta^*)
\Big [{\alpha(1+2c^2_W)\over16\pi s^2_Wc^2_W}(2\ln-\ln^2)\Big ]
\nonumber\\
&&+{3 \alpha^2g_{qR}\sin\theta^*\over 16 s^4_Wc^2_W\mwd (\hat s-m^2_Z)}
[\ln] \Bigg \{(m_i+m_j)\sqrt{\hat s-(m_i-m_j)^2}
\Big [(Z^{N*}_{4i}Z^{N}_{4j}-Z^{N}_{4i}Z^{N*}_{4j})
{m^2_t\over \sin^2\beta }
\nonumber \\
&& -(Z^{N*}_{3i}Z^{N}_{3j}-Z^{N}_{3i}Z^{N*}_{3j})
{m^2_b\over \cos^2\beta} \Big ]
\nonumber\\
&&-
(m_i-m_j)\sqrt{\hat s-(m_i+m_j)^2}
\Big [(Z^{N*}_{4i}Z^{N}_{4j}+Z^{N}_{4i}Z^{N*}_{4j})
{m^2_t\over \sin^2\beta}
-(Z^{N*}_{3i}Z^{N}_{3j}+Z^{N}_{3i}Z^{N*}_{3j})
{m^2_b\over \cos^2\beta}\Big ]\Bigg \}
\nonumber\\
&&
+I^{(3)}_{q}{\alpha^2\bar m_q\bar m_{q'}\sin\theta^*[\ln]
\over 4 m^2_W s^4_W}
\Big \{ {\ln{-\hat t\over\hat s}\over \hat t-M^2_S}
[(m_i+m_j)\sqrt{\hat s-(m_i-m_j)^2}
\nonumber \\
&& - (m_i-m_j)\sqrt{\hat s-(m_i+m_j)^2}]
(Z^{N}_{q'j}Z^{N*}_{qi}+Z^{N*}_{q'i}Z^{N}_{qj})
\nonumber\\
&&- {\ln{-\hat u\over\hat s}\over \hat u-M^2_S}
[(m_i+m_j)\sqrt{\hat s-(m_i-m_j)^2}
-(m_i-m_j)\sqrt{\hat s-(m_i+m_j)^2}]
(Z^{N}_{q'i}Z^{N*}_{qj}+Z^{N}_{q'j}Z^{N*}_{qi})
\Big \}
\nonumber\\
&&
-{Q_q\tilde \beta'_0 \alpha^2 \sin\theta^*\over \cw^4 (\hat s-m^2_Z)}
[(N^{ij}-N^{ij*})(m_i+m_j)\sqrt{\hat s-(m_i-m_j)^2}
\nonumber \\
&& +(N^{ij}+N^{ij*})(m_i-m_j)\sqrt{\hat s-(m_i+m_j)^2}][\ln]~~,
\label{LL+-++}\\[0.5cm]
&& F^{ij}_{+-+-}(\theta^*)=F^{ijB}_{+-+-}(\theta^*)
\Big [{\alpha Q^2_q\over4\pi c^2_W}(2\ln-\ln^2)
-{\alpha[\ln]\over4\pi s^2_W \mwd }\Big ({m^2_t\over \sin^2 \beta}\delta_{qt}
+{m^2_b\over\cos^2\beta  }\delta_{qb}\Big )
\nonumber\\
&& -{\alpha_s\over3\pi}\ln{\hat s\over \mw^2} \Big ]
+F^{ijB~s}_{+-+-}(\theta^*)
\Big [{\alpha(1+2c^2_W)\over16\pi s^2_Wc^2_W}(2\ln-\ln^2)\Big ]
\nonumber\\
&& -{3 \alpha^2g_{qR}\sqrt{\hat s}(1+\cos\theta^*)
\over 16 s^4_Wc^2_W\mwd (\hat s-m^2_Z)}
[\ln]\Bigg \{\sqrt{\hat s-(m_i-m_j)^2}
\Big [(Z^{N*}_{4i}Z^{N}_{4j}-Z^{N}_{4i}Z^{N*}_{4j})
{m^2_t\over \sin^2\beta }
\nonumber\\
&& -(Z^{N*}_{3i}Z^{N}_{3j}-Z^{N}_{3i}Z^{N*}_{3j})
{m^2_b\over \cos^2\beta }\Big ]
- \sqrt{\hat s-(m_i+m_j)^2}
\Big [(Z^{N*}_{4i}Z^{N}_{4j}+Z^{N}_{4i}Z^{N*}_{4j})
{m^2_t\over \sin^2\beta }
\nonumber\\
&&-(Z^{N*}_{3i}Z^{N}_{3j}+Z^{N}_{3i}Z^{N*}_{3j})
{m^2_b\over \cos^2\beta }\Big ]\Bigg \}
\nonumber\\
&& -I^{(3)}_{q}{\alpha^2\sqrt{\hat s}\bar m_q\bar m_{q'}
[\ln]\over 4m^2_W s^4_W}
\Bigg \{{(1+\cos\theta^*) \ln{-\hat t\over\hat s}\over \hat t-M^2_S}
[\sqrt{\hat s-(m_i-m_j)^2}-
\sqrt{\hat s-(m_i+m_j)^2}](Z^{N}_{q'j}Z^{N*}_{qi}
+Z^{N*}_{q'i}Z^{N}_{qj})
\nonumber\\
&&+ {(1+\cos\theta^*) \ln{-\hat u\over\hat s}
\over \hat u-M^2_S}[\sqrt{\hat s-(m_i-m_j)^2}+
\sqrt{\hat s-(m_i+m_j)^2}](Z^{N}_{q'i}Z^{N*}_{qj}
+Z^{N}_{q'j}Z^{N*}_{qi}) \Bigg \}
\nonumber\\
&& +{Q_q\tilde \beta'_0 \alpha^2 (1+\cos\theta^*)
\sqrt{\hat s}\over \cw^4 (\hat s-m^2_Z)}
[(N^{ij}-N^{ij*})(\sqrt{\hat s-(m_i-m_j)^2}
\nonumber \\
&& - (N^{ij*}+N^{ij}) \sqrt{\hat s-(m_i+m_j)^2}) ][\ln]~~,
\label{LL+-+-}\\[0.5cm]
&& F^{ij}_{++++}(\theta^*)=F^{ijB}_{++++}(\theta^*)
\Big [\Big ({\alpha (1+26c^2_W)\over288\pi s^2_Wc^2_W}
+{\alpha\over72\pi c^2_W}\delta_{qd}
+{\alpha\over18\pi c^2_W}\delta_{qu}\Big )(2\ln-\ln^2)
\nonumber\\
&& -{\alpha_s\over3\pi}\ln{\hat s\over \mw^2}
-{\alpha[\ln]\over8\pi s^2_W\mwd }\Big ({m^2_t\over \sin^2\beta }\delta_{qt}
+{m^2_b\over \cos^2\beta }\delta_{qb}\Big )
- {\alpha[\ln]\over16\pi s^2_W\mwd }\Big ({m^2_t\over \sin^2\beta }
+{m^2_b\over \cos^2\beta }\Big )(\delta_{qt}+\delta_{qb})\Big ]
\nonumber\\
&&+I^{(3)}_{q}{\bar m_q\alpha^2 \sqrt{\hat s}[\ln^2]\over 4  m_W s^4_W}
[\sqrt{\hat s-(m_i-m_j)^2}-\sqrt{\hat s-(m_i+m_j)^2}]
\Big [{Z^{N*}_{2j}Z^{N*}_{qi}(1+\cos\theta^*)\over \hat t -\tilde m^2(\tilde q_L)}
\nonumber \\
&& +{Z^{N*}_{2i}Z^{N*}_{qj}(1-\cos\theta^*)\over \hat u -\tilde m^2(\tilde q_L)}\Big ]
+I^{(3)}_{q}{\alpha^2\sqrt{\hat s}\bar m_q[\ln]\over 12 m_W s^4_Wc_W}
[\sqrt{\hat s-(m_i-m_j)^2}-
\sqrt{\hat s-(m_i+m_j)^2}]
\nonumber\\
&& \cdot \Big \{{(1+\cos\theta^*)\over \hat t-M^2_S} \Big [
Z^{N*}_{qi}(Z^{N*}_{1j}s_W- 6I^{(3)}_{q}Z^{N*}_{2j}c_W)\ln{-\hat u\over\hat s}
\nonumber\\
&&+
\Big (6c_W(Z^{N*}_{2j}Z^{N*}_{qi}+Z^{N}_{2i}Z^{N}_{qj})
+4s_W(Z^{N*}_{1i}Z^{N*}_{qj}+Z^{N}_{1j}Z^{N}_{qi})\Big )\ln{-\hat t\over\hat s}\Big ]
\nonumber\\
&&+{(1-\cos\theta^*)\over \hat u-M^2_S}\Big [
Z^{N*}_{qj}(Z^{N*}_{1i}s_W-6I^{(3)}_{q} Z^{N*}_{2i}c_W)\ln{-\hat t\over\hat s}
+\nonumber\\
&&
\Big (6c_W(Z^{N*}_{2i}Z^{N*}_{qj}+Z^{N}_{2j}Z^{N}_{qi})
+4s_W(Z^{N*}_{1j}Z^{N*}_{qi}+Z^{N}_{1i}Z^{N}_{qj})\Big )\ln{-\hat u\over\hat s}
\Big ]\Big \} ~~,
\label{LL++++}\\[0.5cm]
&& F^{ij}_{++--}(\theta^*)=F^{ijB}_{++--}(\theta^*)
\Big  [\Big ({\alpha (1+26c^2_W)\over288\pi s^2_Wc^2_W}
+{\alpha\over72\pi c^2_W}\delta_{qd}
+{\alpha\over18\pi c^2_W}\delta_{qu}\Big )(2\ln-\ln^2)
\nonumber \\
&& -{\alpha_s\over3\pi}\ln{\hat s\over \mw^2}
- {\alpha[\ln]\over8\pi s^2_W \mwd}
\Big({m^2_t\over \sin^2\beta} \delta_{qt}
+{m^2_b\over \cos^2\beta } \delta_{qb}\Big )
 - {\alpha[\ln]\over16\pi s^2_W\mwd }\Big ({m^2_t\over \sin^2\beta }
+{m^2_b\over \cos^2\beta }\Big )(\delta_{qt}+\delta_{qb})\Big ]
\nonumber\\
&&- I^{(3)}_{q}{\bar m_q\alpha^2 \sqrt{\hat s}[\ln^2]\over 4  m_W s^4_W}
 [\sqrt{\hat s-(m_i-m_j)^2}
 + \sqrt{\hat s-(m_i+m_j)^2}]
\Big [{Z^{N*}_{2j}Z^{N*}_{qi}(1-\cos\theta^*)\over \hat t -\tilde m^2(\tilde q_L)}
\nonumber \\
&& +{Z^{N*}_{2i}Z^{N*}_{qj}(1+\cos\theta^*)\over \hat u -\tilde m^2(\tilde q_L)}\Big ]
\nonumber \\
&& -I^{(3)}_{q}{\alpha^2\sqrt{\hat s}\bar m_q[\ln]\over 12 m_W s^4_Wc_W}
[\sqrt{\hat s-(m_i-m_j)^2}+ \sqrt{\hat s-(m_i+m_j)^2}]
\nonumber \\
&& \cdot \Big \{{(1-\cos\theta^*)\over \hat t-M^2_S}\Big [
Z^{N*}_{qi}(Z^{N*}_{1j}s_W-6I^{(3)}_{q} Z^{N*}_{2j}c_W)\ln{-\hat u\over\hat s}
\nonumber\\
&&+
\Big (6c_W(Z^{N*}_{2j}Z^{N*}_{qi}+Z^{N}_{2i}Z^{N}_{qj})
+4s_W(Z^{N*}_{1i}Z^{N*}_{qj}+Z^{N}_{1j}Z^{N}_{qi})\Big )\ln{-\hat t\over\hat s}\Big ]
\nonumber\\
&&+{(1+\cos\theta^*)\over \hat u-M^2_S}\Big [
Z^{N*}_{qj}(Z^{N*}_{1i}s_W-6 I^{(3)}_{q}Z^{N*}_{2i}c_W)\ln{-\hat t\over\hat s}
+\nonumber\\
&& \Big (6c_W(Z^{N*}_{2i}Z^{N*}_{qj}+Z^{N}_{2j}Z^{N}_{qi})
+4s_W(Z^{N*}_{1j}Z^{N*}_{qi}+Z^{N}_{1i}Z^{N}_{qj})\Big )\ln{-\hat u\over\hat s}
\Big ] \Big \} ~~, \label{LL++--}\\[0.5cm]
&& F^{ij}_{+++-}(\theta^*)=F^{ijB}_{+++-}(\theta^*)
\Big [\Big ({\alpha (1+26c^2_W)\over288\pi s^2_Wc^2_W}
+{\alpha\over72\pi c^2_W}\delta_{qd}
+{\alpha\over18\pi c^2_W}\delta_{qu}\Big )(2\ln-\ln^2)
\nonumber \\
&& -{\alpha_s\over3\pi}\ln{\hat s\over \mw^2}
-{\alpha[\ln]\over 8\pi s^2_W\mwd }
\Big ({m^2_t\over \sin^2\beta }\delta_{qt}
+{m^2_b\over \cos^2\beta }\delta_{qb}\Big )
-{\alpha[\ln]\over16\pi s^2_W\mwd }\Big ({m^2_t\over \sin^2\beta }
+{m^2_b\over \cos^2\beta }\Big )(\delta_{qt}+\delta_{qb})\Big ]
\nonumber\\
&&+I^{(3)}_{q} {\bar m_q\alpha^2 [\ln^2]\sin\theta^*\over 4  m_W s^4_W}
[(m_i+m_j)\sqrt{\hat s-(m_i-m_j)^2}
\nonumber \\
&&  + (m_i-m_j)\sqrt{\hat s-(m_i+m_j)^2}]
\Big [{Z^{N*}_{2j}Z^{N*}_{qi}\over \hat t -\tilde m^2(\tilde q_L)}
-{Z^{N*}_{2i}Z^{N*}_{qj}\over \hat u -\tilde m^2(\tilde q_L)}\Big ]
\nonumber\\
&&
+I^{(3)}_{q}{\alpha^2\bar m_q[\ln]\sin\theta^*\over 12 m_W s^4_Wc_W}
[(m_i+m_j)\sqrt{\hat s-(m_i-m_j)^2}
\nonumber\\
&& + (m_i-m_j)\sqrt{\hat s-(m_i+m_j)^2}]
\Big \{ {1\over \hat t-M^2_S}\Big [Z^{N*}_{qi}(Z^{N*}_{1j}s_W-6I^{(3)}_{q}Z^{N*}_{2j}c_W)
\ln{-\hat u\over\hat s}
\nonumber\\
&&+
\Big (6c_W(Z^{N*}_{2j}Z^{N*}_{qi}+Z^{N}_{2i}Z^{N}_{qj})
+4s_W(Z^{N*}_{1i}Z^{N*}_{qj}+Z^{N}_{1j}Z^{N}_{qi})\Big )\ln{-\hat t\over\hat s}\Big ]
\nonumber\\
&&-{1\over \hat u-M^2_S}\Big [Z^{N*}_{qj}(Z^{N*}_{1i}s_W
- 6I^{(3)}_{q}Z^{N*}_{2i}c_W) \ln{-\hat t\over\hat s}
+\nonumber\\
&&
\Big (6c_W(Z^{N*}_{2i}Z^{N*}_{qj}+Z^{N}_{2j}Z^{N}_{qi})
+4s_W(Z^{N*}_{1j}Z^{N*}_{qi}+Z^{N}_{1i}Z^{N}_{qj})\Big )\ln{-\hat u\over\hat s}
\Big ] \Big \} ~~, \label{LL+++-}
\eqa
expressed in terms of the Born amplitudes
$F^{ijB}_{\lambda_1\lambda_2\tau_i\tau_j}$ and its s-channel part
$F^{ijBs}_{\lambda_1\lambda_2\tau_i\tau_j}$ appearing \eg in  (\ref{born},
\ref{Born-s-basic1}-\ref{Born-u-basic}).

\appendix
\newpage

\noindent
{\Large \bf Appendix B}

\renewcommand{\theequation}{B.\arabic{equation}}
\renewcommand{\thesection}{B.\arabic{section}}
\setcounter{equation}{0}
\setcounter{section}{0}

\section{ Parton model kinematics for $\tchi^0_i\tchi^0_j$ production.}

The basic parton model expression for ~the hadron-hadron
collision $A(q_1)B(q_2)\to \tchi^0_i(p_i)+\tchi^0_j(p_j) ...$,   is
\bqa
&&d \sigma(A B \to \tchi^0_i\tchi^0_j  ...)=\nonumber \\
&& \sum_{q_1 q_2} \int\!\int
dx_adx_b~f_{q_1/A}(x_a,Q)f_{q_2/B}(x_b,Q) d \hat{\sigma}
(q_1 q_2 \to \tchi^0_i+\tchi^0_j) ~~ , \label{parton}
\eqa
with $\tchi^0_i,~ \tchi^0_j$ being the two produced massive particles
of mass $m_i$, $m_j$.
Here $f_{q_1/A}(x_a,Q)$ is the distribution function of partons of type
$(q_1 =g,\, q,\, \bar q )$, in the hadron of type $A$ at
a factorization  scale $Q$.\par

Taking the $AB$-c.m. system as the lab-system,
the lab-momenta  of the produced $\tchi^0_i$ and $\tchi^0_j$
are
\bq
p^{\mu}_i=(E_i,p_T,p_i\cos\theta_i)~~~~,~~~
p^{\mu}_j=(E_j,-p_T,p_j\cos\theta_j)~~,
\eq
where  their  transverse momenta are obviously  just opposite
\bq
p_T\equiv p_{Ti}=-p_{Tj}~~,
\eq
 while  their transverse energies $E_{Ti}=\sqrt{p^2_T+m_i^2}$,
$E_{Tj}=\sqrt{p^2_T+m_j^2}$
are  used to define
\bqa
&& x_{Ti}={2E_{Ti}\over\sqrt{s}}, ~~~~~ \beta_{Ti}=p_T/E_{Ti}
=\sqrt{1-{4m_i^2\over s
x^2_{Ti}}} ~, \nonumber \\
&& x_{Tj}={2E_{Tj}\over\sqrt{s}}, ~~~~~ \beta_{Tj}=p_T/E_{Tj}
=\sqrt{1-{4m_j^2\over s
x^2_{Tj}}} \ . \label{pT}
\eqa
Note that
\bq
 E^2_{Tj}=E^2_{Ti}+m^2_j-m^2_i~~~~~~~x^2_{Tj}=
x^2_{Ti}+{4(m^2_j-m^2_i)\over s} \label{ET2-xT2-formulae}
\eq

The rapidities and  production angles of  $\tchi^0_i,~ \tchi^0_j$,
in the lab-system, are related to their energies and
momenta along the beam-axis of hadron $A$, (taken as the
$\hat z$-axis) by
\bq
e^{2y_i} =  {E_i+p_i\cos\theta_i\over E_i-p_i\cos\theta_i} ~~~, ~~~
e^{2y_j} =  {E_j+p_j\cos\theta_j\over E_j-p_j\cos\theta_j}
\eq
The center-of-mass rapidity $\bar y$ of the
$\tchi^0_i\tchi^0_j$ pair, and their respective  rapidities $y_i^*$ in
their own c.m. frame, are defined as
\bq
y_i=\bar{y}+y^*_i ~~~,~~~ y_j=\bar{y}+y^*_j  \  ,
\label{cm-rapidity}
\eq
\bq
\Delta y \equiv y_i-y_j = y_i^*-y_j^* \ \ .
\eq

The fractional  momenta of the
the incoming partons are expressed in terms of their
lab-momenta by (compare (\ref{qq-process}, \ref{gg-process}))
\bqa
q_1= \frac{s}{2}(x_a,0,0,x_a) &~,~ & q_2=\frac{s}{2}(x_b,0,0,-x_b)
~~,~~ q=q_1+q_2~~, \nonumber \\
 q^0={\sqrt{s}\over2}(x_a+x_b)=E_i+E_j &~,~&
q^3={\sqrt{s}\over2}(x_a-x_b)=(p_i\cos\theta_i+p_j\cos\theta_j)~,
\eqa
which lead to
\bqa
x_a & = & {1\over2}[ x_{Ti}e^{y_i}+x_{Tj}e^{y_j}]=\frac{M}{\sqrt
s}\,  e^{\bar y}  \ , \nonumber\\
x_b& =& {1\over2}[x_{Ti}e^{-y_i}+ x_{Tj} e^{-y_j}]= \frac{M}{\sqrt
s} \, e^{-\bar y} \ ,
\label{xab}
\eqa
\bq
\label{shat}
\hat{s}\equiv M^2=(q_1+q_2)^2= x_a x_b s=
{s\over4}[x^2_{Ti}+x^2_{Tj}+2x_{Ti}x_{Tj}\cosh(\Delta y)]\ .
\eq
Using this,  $\hat s, ~x_a, ~x_b$ may be calculated
in terms of the final particle rapidities $y_i,~y_j$ and their
transverse momenta. From them,
$\bar y$ is also obtained, and $(y_i^*,~y_j^*)$ through (\ref{cm-rapidity}).

The remaining  Mandelstam invariants of the subprocesses  satisfy
\bqa
&& \hat{t}=(p_i-q_1)^2 =
m^2_i-M(E^*_i-p^*\cos\theta^*)=m^2_i-\frac{x_{Ti}}{2} M \sqrt{s} e^{-y_i^*}
=m^2_i-{s\over2}x_ax_{Ti}e^{-y_i}
 \nonumber \\
&& = m^2_j-M(E^*_j-p^*\cos\theta^*)=m^2_j-\frac{x_{Tj}}{2} M \sqrt{s} e^{y_j^*}
=m^2_j-{s\over2}x_b x_{Tj}e^{y_j} \ , \label{that}\\
&& \hat{u}=(p_j-q_1)^2
 = m^2_i-M(E^*_i+p^*\cos\theta^*)=m^2_i-{x_{Ti}\over2}M\sqrt{s} e^{y_i^*}
 =m^2_i-{s\over2}x_b x_{Ti}e^{y_i}
\nonumber \\
&& =  m^2_j-M(E^*_j+p^*\cos\theta^*)=m^2_j-{x_{Tj}\over2}M\sqrt{s} e^{-y_j^*}
=m^2_j-{s\over2}x_ax_{Tj}e^{-y_j} \ ,
\label{uhat}
\eqa
\bq
\tau~=~ \frac{\hat s}{s}~=~ x_ax_b  \  \ ,
\eq
where $\theta^*$   describes  $\tchi^0_i$ production angle   in the
$\tchi^0_i\tchi^0_j$-c.m. frame (the $\tchi^0_j$ one being  $\pi-\theta^*$).

The  energies  of the two final particles in their c.m.-frame  are
\bq
E_i^*=\frac{\hat s+m_i^2-m_j^2}{2\sqrt{\hat s}}~~~,~~~
E_j^*=\frac{\hat s+m_j^2-m_i^2}{2\sqrt{\hat s}}~~, \label{Ejstar}
\eq
their  momentum is
\bq
p^*={1\over2M}[(M^2-m^2_i-m^2_j)^2-4m^2_im^2_j]^{1\over2}~~, \label{pstar}
\eq
and their  velocities
\bqa
&& \beta^*_i=p^*/E_i^* =~{[(M^2-m^2_i-m^2_j)^2-4m^2_im^2_j]^{1\over2}
\over M^2+(m^2_i-m^2_j)} \ , \nonumber \\
&& \beta^*_j=p^*/E_j^* =~{[(M^2-m^2_i-m^2_j)^2-4m^2_im^2_j]^{1\over2}
\over M^2 - (m^2_i-m^2_j)} \ .
\label{betastar}
\eqa

We also have
\bq
 \cos\theta^* ={\tanh y^*_i\over\beta^*_i}
= -~{\tanh y^*_j\over\beta^*_j} ~~~,~~~
\sin\theta^* = \frac{p_T}{p^*} ~~, \label{thetastar}
\eq
\bqa
\chi_i\equiv e^{2y_i^*}&= & {\hat{u}-m_i^2 \over
\hat{t}-m_i^2}= {1+\beta^*_i \cos\theta^*
\over 1- \beta^*_i  \cos\theta^*} ~~ , \nonumber \\
\chi_j\equiv e^{2y_j^*}&= & {\hat{t}-m_j^2 \over
\hat{u}-m_j^2}= {1 - \beta^*_j \cos\theta^*
\over 1 +\beta^*_j \cos\theta^*} ~~ , \label{chi} \ .
\eqa

Note that
\bqa
\beta^*_i\cos\theta^*
&=&\frac{\hat u -\hat t}{\hat u +\hat t}=\frac{\chi_i-1}{\chi_i+1}
~,\nonumber \\
\chi_j &= &{\chi_i(m^2_j-m^2_i)+M^2\over\chi_i M^2+m^2_j-m^2_i} ~~, \\
E_{Ti}&= &{E^*_i\over \cosh~ y^*_i} ~~, \\
p^2_T &= &{(M^2+m^2_i-m^2_j)^2\chi_i-M^2m^2_i(1+\chi_i)^2
\over M^2(1+\chi_i)^2}
\eqa
\bq
x^2_{Ti}={4(M^2+m^2_i-m^2_j)^2\chi_i\over M^2s(1+\chi_i)^2}~~~~~~~~
x^2_{Tj}={4(M^2+m^2_j-m^2_i)^2\chi_j\over M^2s(1+\chi_j)^2}
\eq\\

\section{ The basic  distributions at LHC.}

Using (\ref{parton}) we define
\bq
{d\sigma\over dp^2_T dy_i dy_j} =\tau S_{ij} ~, \label{pT-y1-y2}
\eq
or
\bq
{d\sigma\over dM^2 d\chi_i d\bar y} ={M^2+m^2_i-m^2_j\over
M^2(1+\chi_i)^2}~\tau S_{ij} ~, \label{M-chi1-ybar}
\eq
or
\bq
{d\sigma\over dM^2  d\bar y d p^2_T} = \frac{M}{2 s\sqrt{p^{*2}-p^2_T}}
S_{ij} ~, \label{M-ybar-pT}
\eq
where  $S_{ij}$ describes the contribution to (\ref{parton}) from all
 partons.  Grouping together the gluon-gluon and  $q\bar q$
 parton contributions,  we may write
\bq
S_{ij}  \equiv S^g_{ij} +S^{\rm quark}_{ij} \label{Sij}
\eq
\noindent
with
\bq
S^g_{ij}\equiv g(x_a,Q)g(x_b,Q)\frac{d\hat{\sigma}(gg\to \tchi^0_i\tchi^0_j )}{d \hat
t} \ ,
\label{Sgg}
\eq
\bqa
&& S^{\rm quark}_{ij}  \equiv  \nonumber \\
&& \sum_{q} \left [q(x_a,Q)\bar q(x_b,Q)
{d\hat{\sigma}(q\bar q\to \tchi^0_i\tchi^0_j) \over d\hat{t}}
+\bar q(x_a,Q)q(x_b,Q)
{d\hat{\sigma} (\bar q q\to \tchi^0_i\tchi^0_j)\over d\hat{t}} \right ]~,
\label{Sqqbar}
\eqa
where $q=u,d,s,c,b$.

For the numerical calculations presented here, we use as an example the
MRST2003c code for quark and gluon structure functions \cite{MRST},
taking the factorization scale as
\bq
Q=\frac{E_{Ti}+E_{Tj}}{4}~~. \label{Q-scale}
\eq

As seen from above, the basic quantities needed are
$x_a,~ x_b,~ \hat s \equiv M^2,~\hat t, \hat u$.
In case of (\ref{pT-y1-y2}) these are calculated from
(\ref{xab}, \ref{shat}, \ref{that}, \ref{uhat}).
In case (\ref{M-chi1-ybar}),  Eqs.(\ref{chi}, \ref{cm-rapidity},
 \ref{Ejstar}, \ref{pstar}, \ref{betastar}, \ref{thetastar})
 must also be used.

 Starting from this basic distribution, always assuming $m_i>m_j$, and
imposing the cuts
\bq
|y_i| \leq Y_i~~, ~~ |y_j| \leq Y_j ~~{ \rm with}~~ Y_i\leq Y_j ~,
\eq
(in the numerical applications we take $Y_{i,j}=2$) we get the single
variable distributions defined in the following subsections.

\subsection{The transverse energy and  $p_T$ distributions}
\bq
{d\sigma\over dx_{Ti}}  =  \int dy_i \int dy_j {M^2 x_{Ti}\over2}
\, S_{ij}   ~~~, ~~~
{d\sigma\over dp_{T}^2}  =  \int dy_i \int dy_j {\hat s \over s}
\, S_{ij} \ , \label{pT-distribution}
\eq
where $\shat=M^2$ is determined from (\ref{shat}),  $x_{Tj}$ from
(\ref{ET2-xT2-formulae}) and the integration limits are
\bqa
\label{y2-limits}
y_{jmin}&=& \max \Bigg \{ \ln\left ( {x_{Tj}\over2-x_{Ti} e^{-y_i}}
\right )~;~ -Y_j \Bigg \} \ , \nonumber \\
y_{jmax} & = & \min\Bigg \{\ln \left ({2-x_{Ti} e^{y_i}
\over x_{Tj}}
\right )~;~ Y_j\Bigg \} \ ,
\eqa
\bqa
 && y_{imax}=-y_{imin}= \nonumber \\
&& \min\Bigg \{Y_i~;~
\cosh^{-1}\left (\frac{1}{x_{Ti}}(1+{m^2_i-m^2_j \over s})
\right )~;~ \ln\left ({2-x_{Tj}  e^{-Y_j}\over x_{Ti}}
\right ) \Bigg \} \ . \label{y1-limits-for-pT}
\eqa
The $x_{Ti}$ range in (\ref{pT-distribution}) would be
\bq
  \frac{2 m_i}{\sqrt{s}} \leq x_{Ti}  \leq    1+\frac{m_i^2-m_j^2}{s}  ~~,
\label{xT1-range}
\eq
 determined by the requirement that the middle constraint in
(\ref{y1-limits-for-pT}) is  meaningful.

\subsection{The rapidity distribution}
Since the $y_i$ distribution
has to be  symmetric,  we only consider the case
of $y_i>0$, for calculating the $x_{Ti}$-limits. Then in
\bq
{d\sigma\over d y_i} =\int dx_{Ti} \int   dy_j \frac{M^2x_{Ti}}{2}
S_{ij} \ ,
\eq
the $y_j$-limits are given by  (\ref{y2-limits}),
while the limits for the $x_{Ti}$ integration are
\bqa
x_{Timin}& =& x_{\rm Tmin,exp} > {2m_i\over\sqrt{s}} \ ,
\nonumber \\
x_{Timax}& = & \min \Big
\{{1+{m^2_i-m^2_j\over s}\over \cosh y_i}~;~ {2e^{2Y_j \pm y_i}
-\sqrt{\Delta_1}\over e^{2(Y_j \pm y_i)}-1} ~;~
2e^{ -  y_i}\Big \}
\eqa
with
\bq
\Delta_1=4[e^{2Y_j}+{(m^2_i-m^2_j)\over s}(1-e^{2(Y_j \pm y_i)})]
\ ,
\eq
provided $\pm y_i \geq -Y_j$ and (of course) $y_i>0$.  \\

\subsection{The invariant mass distribution}
Using the c.m. rapidity $\bar{y}$ defined above and (\ref{chi}),
one obtains
\bq
{d\sigma\over dM^2} =\int d\chi_i\int  d\bar{y}\,
{(M^2+m^2_i-m^2_j) \over s(1+\chi_i)^2} S_{ij} \ ,
\eq
where the integration limits are
\bqa
 \bar{y}_{max} & = & \min \Big
\{Y_i-{1\over2}\ln\chi_i;~ Y_j-{1\over2}\ln \left({M^2-\chi_i
(m^2_i-m^2_j)\over M^2\chi_i-m^2_i+m^2_j}\right );
~ \ln({\sqrt{s}\over M})
\Big \} , \nonumber \\
 \bar{y}_{min} & = & \max \Big \{-Y_i-{1\over2}\ln\chi_i ;~
 -Y_j-{1\over2}\ln \left ({M^2-\chi_i
(m^2_i-m^2_j)\over M^2\chi_i-m^2_i+m^2_j}\right );
\nonumber
\\ && ~ -\ln({\sqrt{s}\over M}) \Big \}  ,  \label{ybar} \\
 \chi_{imax} &=& \min\Big
\{{1+\beta^*_i\over 1-\beta^*_i} ~;~
\frac{M^2(s+(m^2_i-m^2_j) e^{-2Y_j})}{M^4e^{-2Y_j}+s(m^2_i-m^2_j)}~;~
\nonumber \\
&& \frac{(m^2_i-m^2_j)(1-e^{2(Y_i+Y_j)})+\sqrt{\Delta_2}}{2M^2}~;~
 \frac{s}{M^2}e^{2Y_i} \Big \} , \nonumber \\
 \chi_{imin} &=& \max\Big
\{{1-\beta^*_i\over 1+\beta^*_i}~;~
\frac{M^4e^{-2Y_j}+s(m^2_i-m^2_j)}{M^2(s+(m^2_i-m^2_j) e^{-2Y_j})}~;~
\nonumber \\
&& \frac{2M^2}{(m^2_i-m^2_j)(1-e^{2(Y_i+Y_j)})+\sqrt{\Delta_2}}~;~
 \frac{M^2}{s}e^{-2Y_i} \Big \} ,
\eqa
with
\bq
\Delta_2 = (m^2_i-m^2_j)^2\left ( e^{2(Y_i+Y_j)}-1 \right )^2+
4 M^4e^{2(Y_i+Y_j)} \ .
\eq\\

\subsection{The angular distribution}
 This is given by
\bq
{d\sigma\over d\chi_i} =\int dM^2 \int   d\bar{y}
\, {(M^2+m^2_i-m^2_j) \over s(1+\chi_i)^2} S_{ij} \ ,
\label{chii}
\eq
where the $\bar y$ integration limits are as in (\ref{ybar}), while
for the $M^2$ integration we have the limits
\bqa
&& M^2_{max} = \min
\Big \{\chi_i s e^{2Y_i}; {s\over\chi_i}e^{2Y_i}; M^2_+;
M^{'2}_+; s \Big \} \ , \nonumber \\
&& M^2_{min} = \max
\Big \{ L_1;L_2,L_3; M^2_-;M^{'2}_-\Big \} \ ,
\eqa
with
\bqa
M^2_{\pm}& =&
{1\over2}[\chi_i(m^2_i-m^2_j+se^{2Y_j})\pm\sqrt{\Delta_3}], \\
\Delta_3& =& \chi_i^2(m^2_i-m^2_j+se^{2Y_j})^2-4s(m^2_i-m^2_j)e^{2Y_j}, \\
M'^2_{\pm}&= &{1\over2\chi_i}[m^2_i-m^2_j+se^{2Y_j}\pm\sqrt{\Delta_3'}]
, \\
\Delta_3'&= & (m^2_i-m^2_j+se^{2Y_j})^2-4s(m^2_i-m^2_j)\chi_i^2e^{2Y_j},\\
 L_1&=&{\chi_i (m^2_i-m^2_j)(e^{2(Y_j+Y_i)}-1)
   \over\chi^2_ie^{2(Y_j+Y_i)}-1},\\
 L_2&=&{\chi_i(m^2_i-m^2_j)(e^{2(Y_j+Y_i)}-1)\over
 e^{2(Y_j+Y_i)}-\chi^2_i},\\
 L_3&=&{1\over2}\ [{4 m^2_i\over 1-\chi^2}-2(m^2_i-m^2_j)+\sqrt{\Delta_4}],\\
 \chi&=&{\chi_i-1\over \chi_i+1},\\
\Delta_4 &=& 16\ [\ {m^4_i \over (1-\chi^2)^2}-\frac{m^2_i(m^2_i-m^2_j)}{1-\chi^2}\ ]
 .
\eqa

%\clearpage

\newpage

\clearpage

\clearpage

\clearpage

\begin{figure}[t]
\vspace*{-2cm}
\[
\hspace{-1.0cm}\epsfig{file=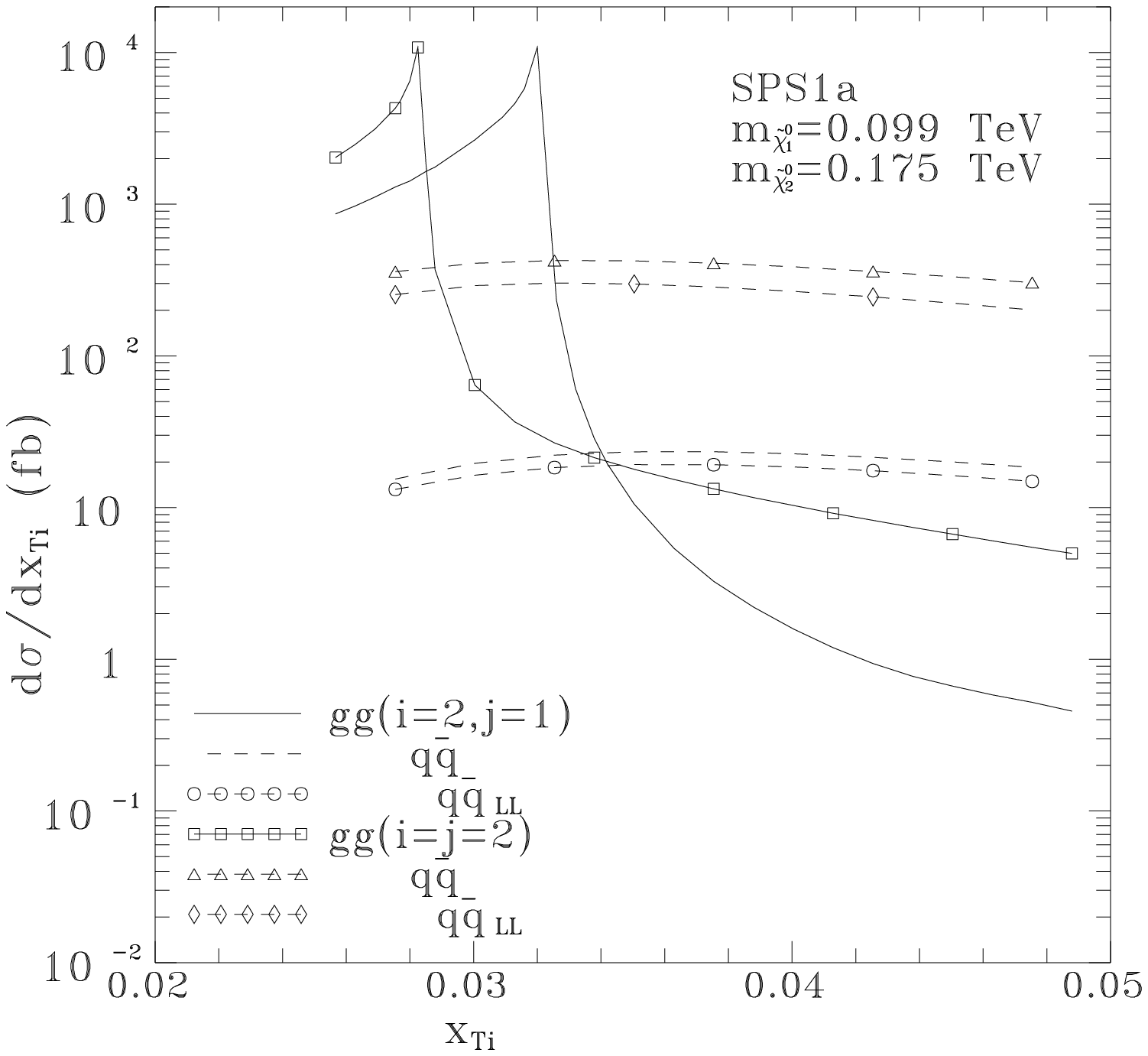,height=9.cm}
\]
\vspace*{-1cm}
\caption[1]{The $x_{Ti}$ distribution in SPS1a  of \cite{Snowmass}.}
\label{xt-SPS1a-fig}
\end{figure}

\begin{figure}[b]
\vspace*{-1cm}
\[
\hspace{-1.0cm}\epsfig{file=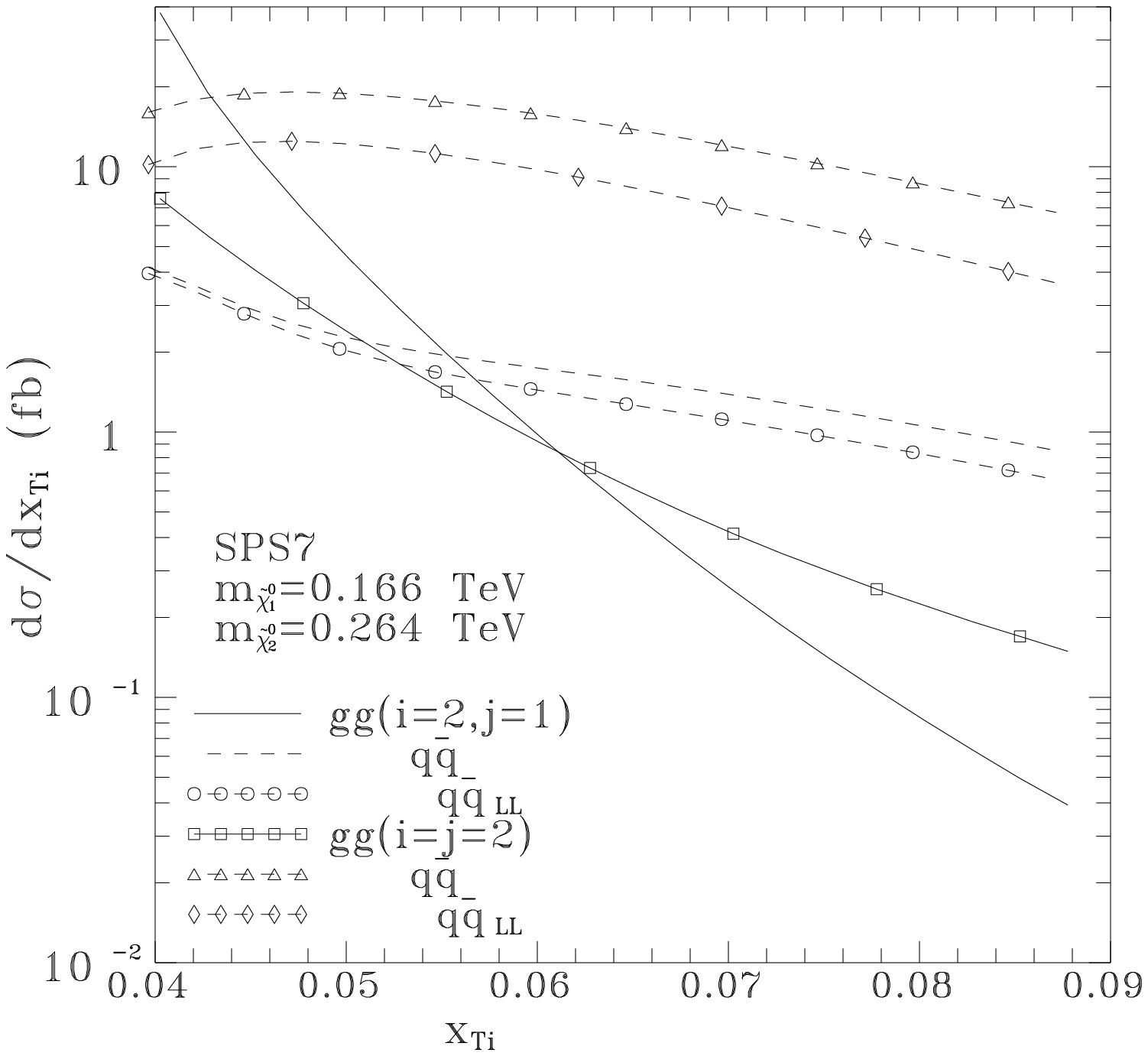,height=9.0cm}
\]
\vspace*{-1cm}
\caption[1]{The $x_{Ti}$ distribution in  SPS7 of \cite{Snowmass}.}
\label{xt-SPS7-fig}
\end{figure}

\clearpage

\begin{figure}[t]
\vspace*{-2cm}
\[
\hspace{-1.cm}\epsfig{file=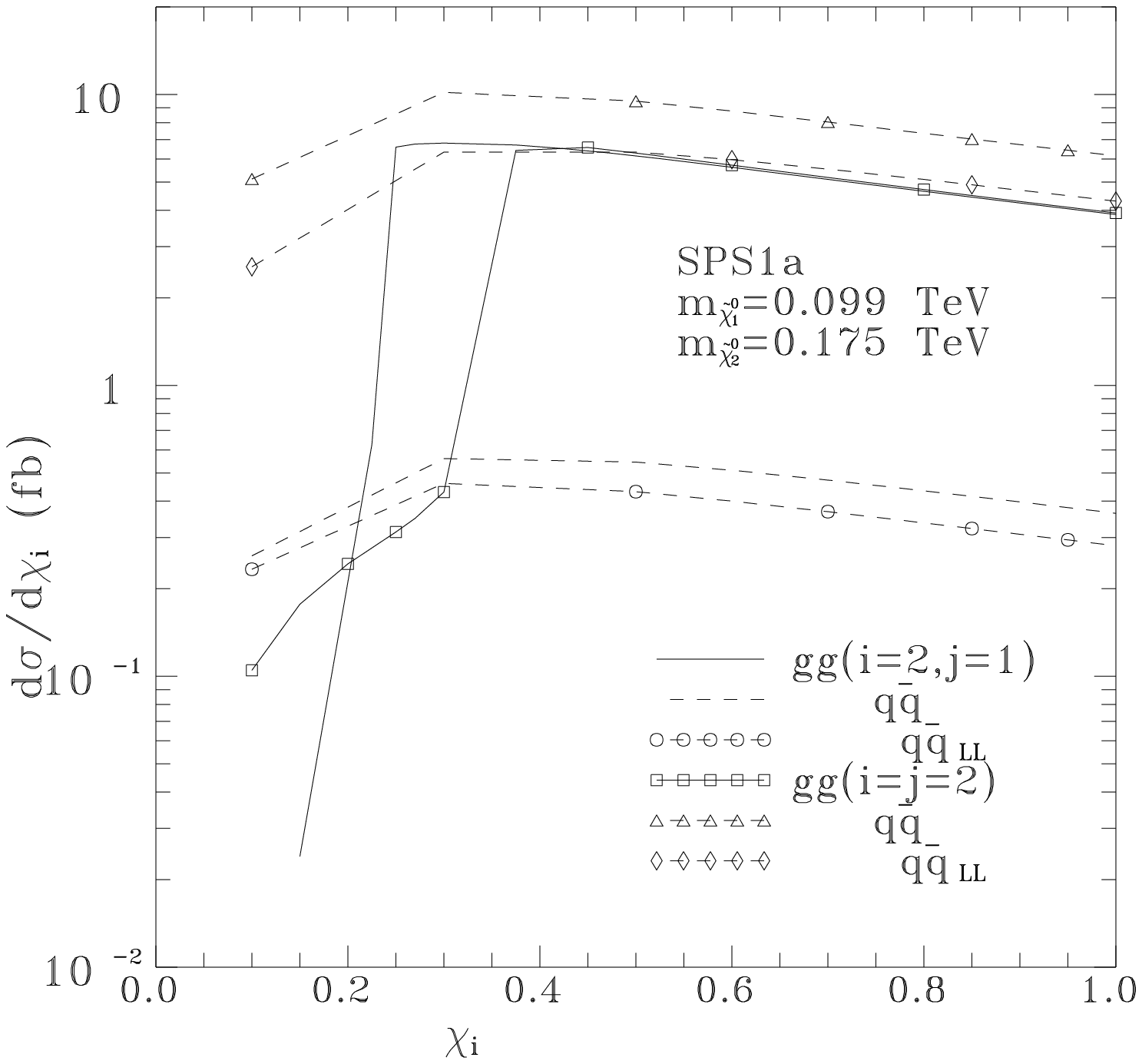,height=9.cm}
\]
\vspace*{-1cm}
\caption[1]{The $\chi_i$ distribution in SPS1a  of \cite{Snowmass}.}
\label{chi-SPS1a-fig}
\end{figure}

\begin{figure}[b]
\vspace*{-1cm}
\[
\hspace{-1.cm}\epsfig{file=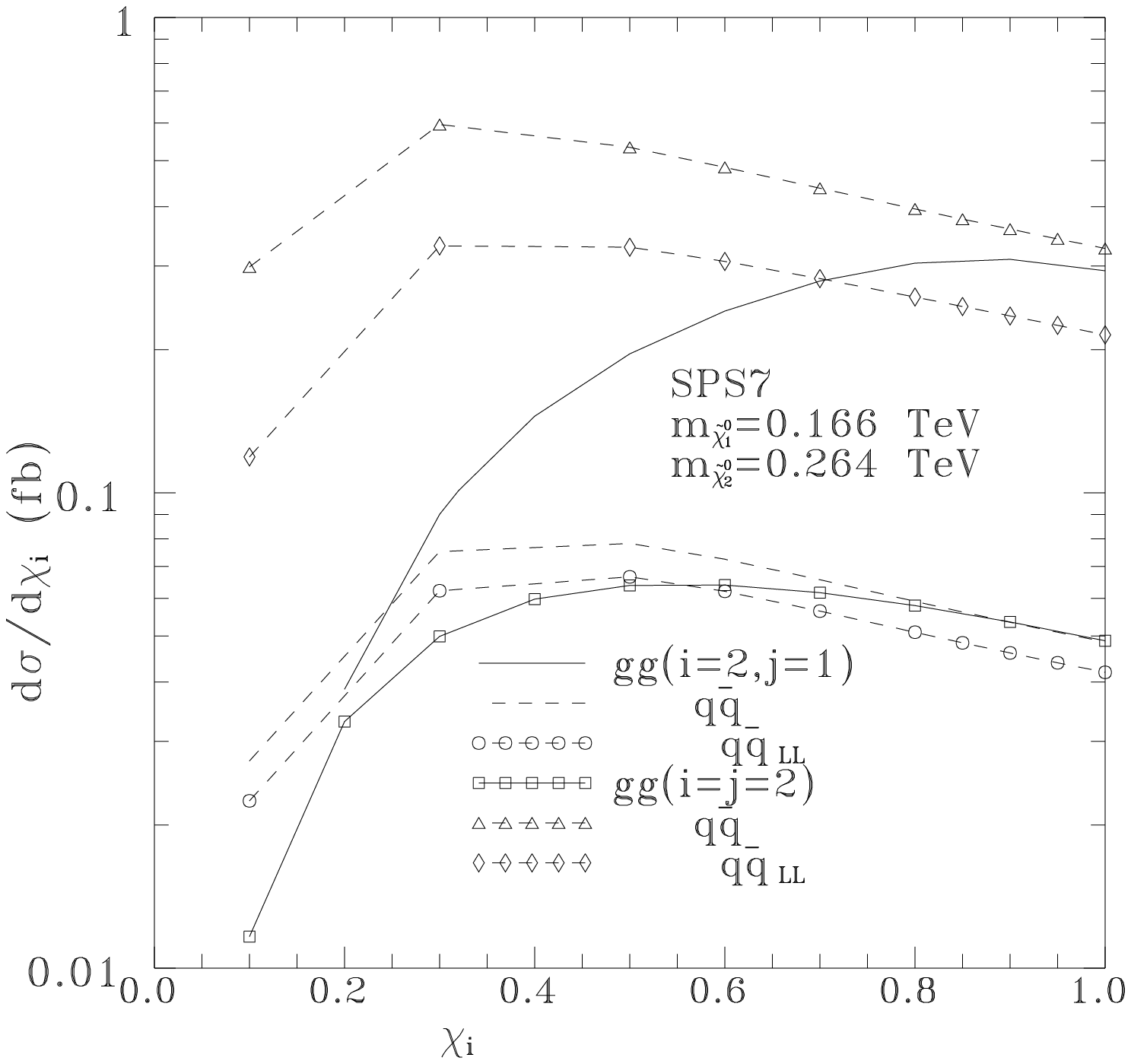,height=9.cm}
\]
\vspace*{-1cm}
\caption[1]{The $\chi_i$ distribution in SPS7 of \cite{Snowmass}.}
\label{chi-SPS7-fig}
\end{figure}


\newpage

\begin{thebibliography}{99}


\bibitem{SUSYsearches1} see e.g. D.P. Roy, \polon{B34}{3417}{2003},
hep-ph/0303106; F.E. Paige, hep-ph/0307342, hep-ph/0211017.
%
\bibitem{SUSYsearches2} B.C. Allanach, G.A. Blair, S. Kraml,
H.U. Martyn, G. Polesello,  W. Porod and  P.M. Zerwas,
hep-ph/0403133; K. Desch, J. Kalinowski, G. Moortgat-Pick,
M.M. Nojiri and G. Polesello, hep-ph/0312069; G. Weiglein, hep-ph/0404108.
%
\bibitem{LSP} A.H. Chamseddine, R. Arnowitt and P. Nath,
\prl{49}{970}{1982};
R. Barbieri, S. Ferrara and C.A. Savoy, \pl{B119}{343}{1982};
L. Hall, J. Lykken and S. Weinberg \pr{D27}{2359}{1983};
L. Randall and R. Sundrum, \np{B557}{79}{1999};
G. Giudice, M. Luty, H. Murayama and R. Rattazzi,
\jhep{9812}{027}{1998};
J.A. Bagger, T. Moroi and E. Poppitz, \jhep{0004}{009}{2000}.
%
\bibitem{DMLSP}  M. Drees, Pramana {\bf 51}:87(1998);
M.S. Turner, J.A. Tyson, astro-ph/9901113, \rmp{71S}{145}{1999};
M.M. Nojiri, hep-ph/0305192; M. Drees hep-ph/0210142;
J. Ellis, astro-ph/0304183; D.N. Spergel \etal arXiv:astro-ph/0302209;
G. Jungman, M. Kamionkowski and K. Griest,
\prep{267}{195}{1996}; M. Kamionkowski, hep-ph/0210370;
L. Roszkowski,  hep-ph/0404052.
%
\bibitem{DMobs}D.S.Akerib, S.M. Carrol, M. Kamionkowski and S. Ritz,
hep-ph/0201178.
%
\bibitem{DMann} G.J. Gounaris, J. Layssac, P.I. Porfyriadis,
F.M. Renard, arXiv:hep-ph/0309032, to appear in Phys.Rev.D.
%
\bibitem{gammagammaDM} G.J. Gounaris, J. Layssac, P.I. Porfyriadis,
F.M. Renard, arXiv:hep-ph/0311076, EPJ {\bf C32}:561(2004).
%
\bibitem{Snowmass}
B.C. Allanach et al, \epj{C25}{113}{2002}, hep-ph/0202233;
G. Weiglein, hep-ph/0301111.
%
\bibitem{Arnowitt}
R. Arnowitt and B. Dutta, talk at 10th Int. Conf. on Supersymmetry and
Unification of Fundamental Interactions (SUSY02), Hamburg, 2002
(hep-ph/0211042).
%
\bibitem{CDG}C.H. Chen, M. Drees and J.F. Gunion, \pr{D55}{330}{1997}
and (E) \pr{D60}{039901}{1999}; J. Amundson \etal, Report of the Snowmass
Sypersymmetry Theory Working Group, hep-ph/9609374;
A. Djouadi, Y. Mambrini and M. M\"uhlleitner, hep-ph/0104115,
\epj{C20}{563}{2001}.
%
\bibitem{MRST} MRST2003c.f can be obtained from
http://durpdg.dur.ac.uk/hepdata/pdf.html.
See also A.D. Martin, R.G. Roberts, W.J. Stirling and R.S.
Thorne, hep-ph/0307263; R.S. Thorne, hep-ph/0309343.
%
\bibitem{Dawson}S. Dawson , E. Eichten and C. Quigg, \pr{D31}{1581}{1985}.
%
\bibitem{Baer1}H. Baer, K. Hagiwara and X. Tata, \pr{D35}{1598}{1987}
%
\bibitem{BRV}
M. Beccaria, M. Melles, F.M. Renard, S. Trimarchi and C. Verzegnassi,
arXiv:hep-ph/0304110, IJMP {\bf A18}:5069 (2003).
%
\bibitem{Denner} A. Denner and S. Pozzorini \epj{C18}{461}{2001}; ibid
 \epj{C21}{63}{2001}.
%
\bibitem{qqLHC} M. Beccaria, F.M. Renard and C. Verzegnassi,
hep-ph/0402028, to appear in Phys. Rev. D.
%
\bibitem{Beenakker} W. Beenakker, M. Klasen, M Kr\"amer .Plehn,
M. Spira and P.M. Zerwas, \prl{83}{3780}{1999}.
%
\bibitem{plato} PLATON codes can be downloaded
from http://dtp.physics.auth.gr/platon/
%
\bibitem{JacobW}M. Jacob and G.C. Wick, \aop{7}{404}{1959}.
%
\bibitem{Liang}H.Liang, Ma Wen-Gan, Jiang Yi, Zhou Mian-Lai and Zhou
Hong, Commun. Theor. Phys. {\bf 34}:115 (2000).
%
\bibitem{LeMouel}see \eg G.J. Gounaris, C. Le Mou\"el and
P.I. Porfyriadis \pr{D65}{035002}{2002};
G.J. Gounaris and C. Le Mou\"el  \pr{D66}{055007}{2002}.
%
\bibitem{Rosiek} J. Rosiek, \pr{D41}{3464}{1990}, hep-ph/9511250(E).
%
\bibitem{epem}H. Baer, R. Munroe,  X. Tata, \pr{D54}{6735}{1996};
H. Baer, A. Bartl, D. Karatas, W. Majerotto, X. Tata \ijmp{A4}{4111}{1989}.





\end{thebibliography}
\end{document}